\documentclass{aa}
\usepackage{aas_macros}
\usepackage{graphicx}
\usepackage{booktabs}
\usepackage{multirow}
\usepackage{physics}
\usepackage[table,xcdraw]{xcolor}
\usepackage[version=3]{mhchem}
\usepackage{colortbl}
\usepackage{txfonts}
\usepackage{subcaption}
\usepackage{textcomp}
\usepackage[export]{adjustbox} 
\usepackage{pdflscape}

\usepackage[colorlinks,citecolor=blue,linkcolor=red,urlcolor=cyan]{hyperref}
\newcommand{\tess}{TESS}

\begin{document} 
%author list is not yet final

   \title{Rapidly oscillating Ap stars observed with TESS}
   \subtitle{The LAMOST Ap sample and 49 Cam}

   \author{Inês Rolo %ORCID(0009-0002-9173-9734)
          \inst{1,2}
          \and
          Daniel L. Holdsworth \inst{3,4} %ORCID(0000-0003-2002-896X)
          \and
          Margarida S. Cunha \inst{1} %ORCID(0000-0001-8237-7343)
          \and
          Victoria Antoci \inst{5} %ORCID(0000-0002-0865-3650)
          \and
          Donald W. Kurtz \inst{6,7} %ORCID(0000-0002-1015-3268)
          \and
          Rahul Jayaraman\inst{8} %ORCID(0000-0002-7778-3117)
          \and
          Ângela R. G. Santos \inst{1,2,9} %ORCID(0000-0001-7195-6542)
          }

   \institute{Instituto de Astrofísica e Ciências do Espaço, Universidade do Porto, CAUP, Rua das Estrelas, 4150-762 Porto, Portugal (\email{ines.rolo@astro.up.pt})
         \and
             Departamento de Física e Astronomia, Faculdade de Ciências, Universidade do Porto, Rua do Campo Alegre 687, 4169-007 Porto, Portugal
        \and South African Astronomical Observatory, PO Box 9, Observatory, 7935, Cape Town, South Africa
        \and School of Physics, Engineering and Technology, University of York, Heslington, York, YO10 5DD, UK 
        \and DTU Space, Technical University of Denmark, Elektrovej 327, Kgs., Lyngby 2800, Denmark
        \and Centre for Space Research, North-West University, Dr Albert Luthuli Drive, Mahikeng 2735, South Africa
        \and Jeremiah Horrocks Institute, University of Lancashire, Preston PR1 2HE, UK
        \and Department of Astronomy, Cornell University, 122 Sciences Drive, Ithaca, NY 14853, USA
        \and Universit\'e Paris-Saclay, Universit\'e Paris Cit\'e, CEA, CNRS, AIM, 91191 Gif-Sur-Yvette, France}
        
   \date{Received;Accepted}

% \abstract{}{}{}{}{} 
% 5 {} token are mandatory
 
  \abstract
  % context heading (optional)
  % {} leave it empty if necessary  
   {The rapidly oscillating chemically peculiar A-type (roAp) stars offer valuable insights into the internal physical processes of all stars, but their study is challenged by their rarity. The large-scale Transiting Exoplanet Survey Satellite (\tess) and \textit{Gaia} surveys have allowed for the collection of data for a sizeable dataset of roAp stars. Nevertheless, asteroseismic data obtained with \tess\ has not been explored to its full potential.}
  % aims heading (mandatory)
   {We develop an algorithm capable of analysing large quantities of \tess\ data to search for new roAp stars and increase the current sample. We focus on data products that have not been previously explored for the search of roAp stars, namely the \tess\ 200-s Full Frame Images (FFIs) and 20-s cadence light curves.}
  % methods heading (mandatory)
   {20-s and 200-s cadence light curves of target stars are retrieved from the Mikulski Archive for Space Telescopes (MAST) server and cleaned. Discrete Fourier Transforms (DFTs) are computed for each light curve which are used to pre-whiten the data to remove any low frequency signals. A final DFT is calculated which is used to classify stars as non-pulsating (NP), $\delta$ Scuti or roAp based on the remaining signal.}
  % results heading (mandatory)
   {We apply our algorithm to two independent datasets: i) $\sim$2700 Ap stars spectroscopically classified with the Large Sky Area Multi-Object Fiber Spectroscopic Telescope (LAMOST) and observed by TESS in the 200-s Full Frame Images (FFIs) and {\sc{ii}}) all TESS 20-s cadence light curves available for known or candidate roAp stars. These two samples have no overlap, i.e. none of the LAMOST stars have been observed with 20-s cadence. We identify four new roAp stars: TIC 312111544, TIC 252881095, TIC 46054683, and 49 Cam (TIC 393276640). We find evidence in TESS data that TIC 252881095 may be part of a binary system. If the tentative $\sim$30-d orbital signal is confirmed, TIC\,252881095 could be one of the shortest-period roAp binary currently known. Furthermore, the detection of high-frequency pulsations in 49 Cam is particularly relevant, as this well-known roAp candidate star is here confirmed to be roAp based on TESS 20-s cadence data.}
  % conclusions heading (optional), leave it empty if necessary 
   {}


   \maketitle
%
%-------------------------------------------------------------------

\section{Introduction}

Chemically peculiar, highly magnetic A-type (Ap) stars have long been of scientific interest \citep[e.g.][]{1958Babcock,1967Wolff,1982Landstreet,2003Cunha,2007Auriere,2018Mathys,2019Sikora,2020Mathys,2021Deal,2024Kochukhov}. The combination of slow rotation, the lack of significant stellar winds and the absence of a large convective outer layer make for a remarkably stable atmosphere. This environment potentiates the efficiency of gravitational settling and radiative levitation, i.e. atomic diffusion, at their surface resulting in strong chemical anomalies accessible to observations \citep[e.g.][]{2001Shavrina,2011Oleg,Joshi2016,2015Leblanc,2017Khalack,2018Ndiaye,2022Folsom}. 

Typically, Ap stars are characterised by a strong, large-scale, dipolar magnetic field \citep[e.g.][]{2018Kochukhov} that impacts the diffusion rate according to the field geometry \citep{2015Balona} and is thought to suppress convection \cite[e.g.][]{2001Balmforth}. This gives rise to the accumulation of chemical elements, most notably rare earth elements, observed as spots at the stellar surface \citep[e.g.][]{2001Shavrina,2011Oleg,Joshi2016,2015Leblanc,2016Bychkhov,2017Khalack,2018Ndiaye,2019Sikora,2020Mathys,2022Folsom}. These chemical spots remain stable for several decades on account of the stability of the magnetic field, modulating the observed flux and allowing for the precise determination of the stellar rotation period \citep[e.g.][]{2016Bychkhov,2019Sikora,2020Mathys}. Thus, Ap stars offer a unique opportunity to better understand the processes behind the transport of chemical elements as well as how the magnetic field influences them. This remains a fundamental problem in stellar modelling \citep[e.g.][]{1945Cowling,2017Braithwaite,2019Keszthelyi,2019Cantiello,2020Schneider}.

About 5.5\% of all known Ap stars \citep{Cycle_2_2024} show short period (4.7-23.6 minutes), low-degree ($l\lesssim$ 3), high radial-order ($n\gtrsim10$) pulsations \citep[e.g.][]{Kurtz1982,2012Alentiev,2019Cunha,Jayaraman2021,Cycle_1_2021,Cycle_2_2024}.
These rare pulsators, first discovered by \citet{1978Kurtz,Kurtz1982}, are known as the rapidly oscillating Ap (roAp) stars. These stars can be found in the main sequence, sharing the base of the classical instability strip with $\delta$\,Scuti, $\gamma$ Doradus and hump and spike stars \citep[e.g.][]{2000Breger,2002Cunha,Aerts2010,chaplin2017asteroseismic,2019Antoci,2025Antoci} and spanning an effective temperature range of $\sim$ 6000\,K to 9000\,K \citep{2019Cunha,2021Deal}. 

The oscillations of roAp stars are, predominantly, acoustic p modes that are thought to be driven by the opacity ($\kappa$) mechanism, which acts as a heat engine \citep[e.g.][]{1996Gunter,1999Cunha,Aerts2010}. In ionization zones (e.g. hydrogen or helium) compression leads to radiation being temporarily trapped which causes pressure to build up. This increased pressure forces the layer to expand and cool. The $\kappa$-mechanism provides energy to pulsations in regions where opacity increases compared to adjacent layers during compression. In pulsating Ap stars specifically, the strong magnetic field is thought to suppress convective energy transport in the outer envelope \citep[e.g.][]{2001Balmforth,2002Cunha,2005Saio,2010Saio}. This has two key effects: (1) it modifies the temperature gradient, and (2) it enhances helium settling, leading to a reduced helium abundance in the He-\textsc{ii} ionization zone \citep[e.g.][]{2005AandTheado}. These changes, combined with the influence of the Lorentz force on pulsations \citep{2000Cunha,2005Saio}, work against the driving of pulsations by the $\kappa$-mechanism in this region. In particular, acoustic oscillations couple with magnetic waves allowing pulsation energy to leak along magnetic field lines, ultimately suppressing mode excitation \citep{2000Cunha,2001Balmforth,2004Saio,2005Saio,2006Cunha}. As such, the $\kappa$-mechanism only becomes effective in shallow layers in the envelope, particularly in the hydrogen ionization layer where the pulsations of roAp stars are thought to be driven.

Observations challenge this theory as models are not capable of reproducing the highest frequency modes found in some roAp stars \citep{Cunha2013,2018Holdsworth}. \citet{Cunha2013} proposed that a mechanism reliant on turbulent pressure might be responsible for the excitation of these high frequency pulsations. Unlike the $\kappa$-mechanism, the turbulent pressure mechanism relies on the interaction between convection and oscillations. Specifically, this mechanism becomes effective in convectively unstable regions. Here, the non-local, time-dependent convection interacts dynamically with the pulsations: the turbulent motions of convective elements generate pressure fluctuations that can resonantly couple with stellar oscillation modes, transferring energy to them cyclically \citep[e.g.][]{2001Balmforth,Cunha2013}. Although convection is often associated with stochastic excitation, the non-local, time-dependent convection treatment allows turbulent-pressure driving to act coherently, producing stable, phase-coherent pulsations. The energy of such pulsations would then be reflected by the strong magnetic field back to the star \citep[e.g.][]{2008Sousa}. Pulsations driven by turbulent pressure should appear clustered around different frequency regions. This has been observed in some roAp stars that show well separated modes in their oscillation spectra \citep[e.g.][]{2015Medupe,2019Cunha}. This mechanism has also been proposed to explain the high frequencies of $\delta$ Scuti stars \citep{2014Antoci}.

In roAp stars, the magnetic and rotation axes are typically misaligned, while the pulsation axis is closely aligned with the magnetic axis \citep[e.g.][]{1985Shibahashi,2002Bigot,2011Bigot,2019Holdsworth}. Because the Lorentz force associated with the strong global magnetic field dominates over the Coriolis force, the pulsation geometry is set primarily by the magnetic axis \citep[e.g.][]{1990KurtzGoode}. This contrasts with what is observed for most p-mode pulsators where the pulsation axis typically aligns with the rotation axis instead \citep{Kurtz1982,2006Kurtz}. {As a result, when a roAp star rotates, the observer's view of the pulsation changes. In the Fourier domain of an roAp star light curve, this causes splitting of pulsation mode frequencies into sidelobes that appear separated from the central mode frequency by exactly the rotation frequency \citep[e.g.][]{1985Shibahashi,1990KurtzGoode,Kurtz1982,1999Cunha}. \citet{Kurtz1982} developed the oblique pulsator model to explain this phenomenon. Oblique pulsation provides clear constraints on mode geometry. If a mode is non-distorted, i.e. described by a single spherical harmonic, we expect a multiplet of 2$l$+1 components to appear in the frequency domain, where $l$ is the degree of the mode representing the number of nodes appearing on the stellar surface. We note that this oblique-pulsator multiplet is distinct from Ledoux rotational splitting, commonly observed in other non-radial pulsators (e.g.\ $\delta$~Scuti/$\gamma$~Doradus), where the observed frequency splitting arises from the transformation of pulsation frequencies from the co-rotating frame into the observer's frame \citep{1951Ledoux}.

Currently, there are $\sim$ 100 roAp stars known in the literature, identified using both ground-based and space-based data \citep[e.g.][]{1994MartinezKurtz,2013Kochukhov,Paunzen2015,Joshi2016,2019Cunha,Jayaraman2021,Cycle_1_2021,Cycle_2_2024}. A large scale comprehensive study of the characteristics of roAp stars is highly desirable, as their unique asteroseismic characteristics provide valuable information for the development of theoretical models. However, their rarity is a significant obstacle. The {Transiting Exoplanet Survey Satellite \citep[\tess]{2014Ricker}} has, since its launch, greatly aided the unbiased search and discovery of new roAp stars \citep{Cycle_1_2021,Cycle_2_2024}. However, much of the publicly available \tess\ data has yet to be explored in the context of roAp stars. To address this we developed a versatile tool capable of analysing large quantities of data for the detection and classification of roAp stars. The present article is organised as follows: in Section~\ref{data} we describe the data we aim to explore as well as the tool we developed. In Sections \ref{200-s } and \ref{20s} we show the results of this search, performed on two well constrained datasets. And lastly in Section~\ref{conc} we lay out our main conclusions as well as our next steps.

\section{Data and search methodology}
\label{data}
\subsection{Observational data}
Some TESS data products remain underexplored in the context of roAp-star searches, and hold significant potential for expanding and better understanding the current sample of known roAp stars. These are the available 20-s cadence data collected for known and candidate roAp stars, and the data contained in 200-s cadence TESS Full Frame Images (FFIs).

In its nominal mission \tess\ collected data every 120-s for a selection of $\sim$ 20,000 target stars per sector, which allowed for a large-scale, homogeneous search for high-frequency stellar variability \citep{2019Cunha,2019Antoci,Cycle_1_2021,Cycle_2_2024}. The 20-s cadence data, available since the first extended mission, increases the frequency range beyond the 360\,d$^{-1}$ Nyquist limit to 2160 \,d$^{-1}$, making it possible to search for higher-frequency pulsation modes, and harmonics of lower-frequency modes. In addition, the 20-s cadence data may also improve sensitivity at lower-amplitude mode and sidelobe frequencies thanks to the higher signal-to-noise-ratio (SNR). As discussed in \cite{2022Huber}, while light curves with both cadences are reduced by the Science Processing Operations Center (SPOC) pipeline \citep{2016Jenkins}, for the 120-s data, a single cosmic-ray hit triggers the rejection of the entire 120-s exposure resulting in an effective loss of $\sim$20\% of exposures. In contrast, for the 20-s cadence data products, a cosmic ray affects at most a single 20-s exposure. 

Because 20-s cadence observations are only available for a limited number of targets through approved \tess\ proposals, our 20-s sample is restricted to a set of 37 stars: 31 confirmed and 6 candidate roAp stars\footnote{TESS GI Proposal ID: G04112 (PI: Margarida Cunha), \emph{Searching For The Missing Frequencies In roAp Stars}; \url{https://heasarc.gsfc.nasa.gov/docs/tess/data/approved-programs/cycle4/G04112.txt}.}. This cadence allows to look for pulsation frequencies beyond the 120-s Nyquist frequency predicted by the turbulent mixing mechanism which, so far, have not been observed \citep[e.g.][]{Cunha2013,2019Cunha,Cycle_1_2021,Cycle_2_2024}. In this work, we analyse for the first time the 20-s cadence light curves not only to search for these theoretically predicted oscillations, but also to identify any previously undetected pulsation modes, and harmonics of lower frequency modes. 

In the second extended mission \tess\ (Sectors $\geq 56$) started collecting Full Frame Images (FFIs) with a cadence of 200-s, providing sufficient temporal resolution to detect typical roAp pulsations \citep[$\nu> 60$ d$^{-1}$;][]{2019Cunha,Cycle_1_2021,Cycle_2_2024}. These data provide an opportunity to substantially increase our sample of known roAp stars, containing information on millions of targets. The Massachusetts Institute of Technology (MIT) Quick-Look Pipeline \citep[QLP;][]{2020aHuang,2020bHuang} provides processed light curves extracted from \tess\ FFIs for targets with \tess\ magnitudes $<$ 13.5 mag, available through the Mikulski Archive for Space Telescopes\footnote{\url{https://archive.stsci.edu/missions-and-data/tess}} (MAST).

Selecting all stars with QLP light curves from Sector 56 onwards to search for high-frequency pulsations would be a computationally expensive task, with low return when considering the rarity of roAp stars and the number of stars that would be processed. Therefore, we built a target list based on the Ap spectral classifications presented by \citet{2020AandA...640A..40H} and \citet{2023Shi}. These works utilised spectra collected by the Large Sky Area Multi-Object Fiber Spectroscopic Telescope (LAMOST) and searched for the characteristic flux depression that Ap stars show in the 5200\,\AA\ region \cite[e.g.,][]{2018CoSka..48..218M}. This spectral feature is caused predominantly by Fe lines, but with contributions from Si and Cr particularly at lower temperatures \citep{2007AandA...469.1083K}. Both works employed modified versions of the spectral classification code MKCLASS \citep{Gray_2014} that uses lines of Si\,{\sc{ii}}, Sr\,{\sc{ii}}, and Cr\,{\sc{ii}} at 4077\,\AA, the blend of Si\,{\sc{ii}} and Eu\,{\sc{ii}} at 4130\,\AA, and the Eu\,{\sc{ii}} 4205\,\AA\ line to identify Ap stars. \citet{2023Shi} performed a visual inspection of all of the classified spectra, while \citet{2020AandA...640A..40H} performed `spot checks' on 10\% of their sample and found good agreement between manual and automatic classification. While we acknowledge no system is perfect in spectral classification of chemically peculiar stars, we have confidence that there are few false positive identifications in the combined results of \citet{2020AandA...640A..40H} and \citet{2020Shi}, which yield a sample of 2700 Ap stars.

\subsection{Methodology for roAp star search and identification}
\label{alg}

Our tool consists of four steps: data download, pre-whitening, frequency extraction and classification of the pulsating stars. The algorithm collects the time and simple aperture photometry flux from the downloaded light curves and cleans the signal according to the provided \texttt{QUALITY} flags. Only data points without anomaly flags are kept for analysis. Then, a Discrete Fourier Transform (DFT) is calculated, as described in \citet{1985MNRASKurtz}. This first transformation of the signal to the frequency domain serves the important purpose of noise estimation. The low-frequency rotation signal, a consequence of the existence of surface chemical spots, tends to dominate the amplitude spectrum, affecting the apparent noise level of the DFT. As our main goal is to identify high-frequency pulsations ($>$60\,d$^{-1}$), we start by removing all significant peaks in the frequency range 0--10\,d$^{-1}$ (pre-whitening). Our algorithm then determines the noise level of the signal by identifying noise segments in the DFT and selecting the segment that spans the largest frequency range, deemed the most representative of the overall noise (Figure~\ref{fig:noise_win}). The amplitude of the noise level is approximately two times the average value of the noise ($\sigma_{\rm DFT}$) in that segment. We repeat this process twice, as the initial noise-level estimation can be unreliable when the rotation signal dominates the DFT.
\begin{figure}[h!]
    \centering
    \includegraphics[width=\columnwidth]{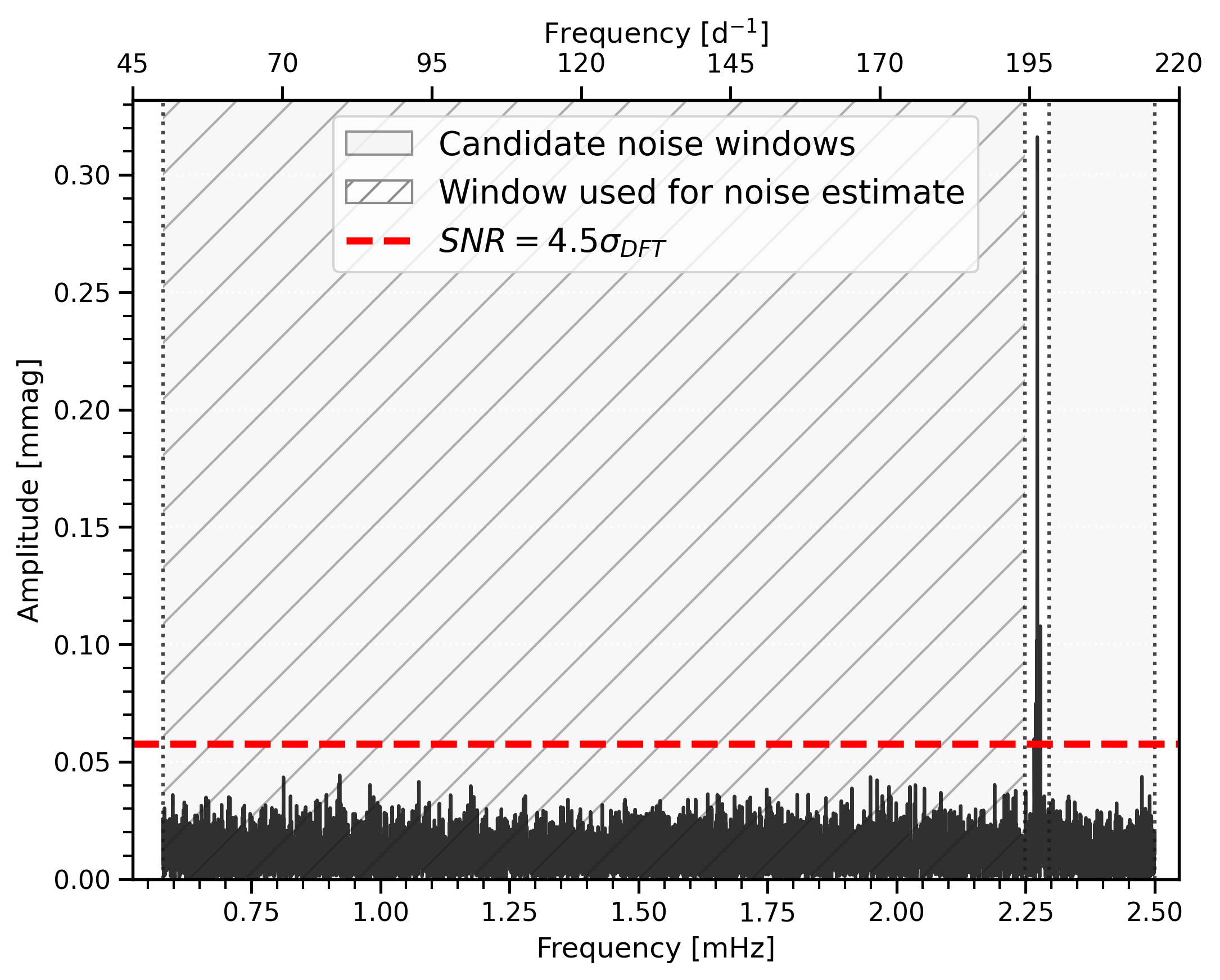}
    \caption{Example of the SNR=$4.5\sigma_{\rm DFT}$ threshold from the amplitude spectrum of the roAp star TIC\,318007796. The light-grey shaded regions indicate the candidate peak-free frequency windows, and the hatched region marks the window adopted to compute the noise level (red dashed line).}
    \label{fig:noise_win}
\end{figure}  
\begin{figure*}[h!]
    \centering
    \includegraphics[width=\textwidth]{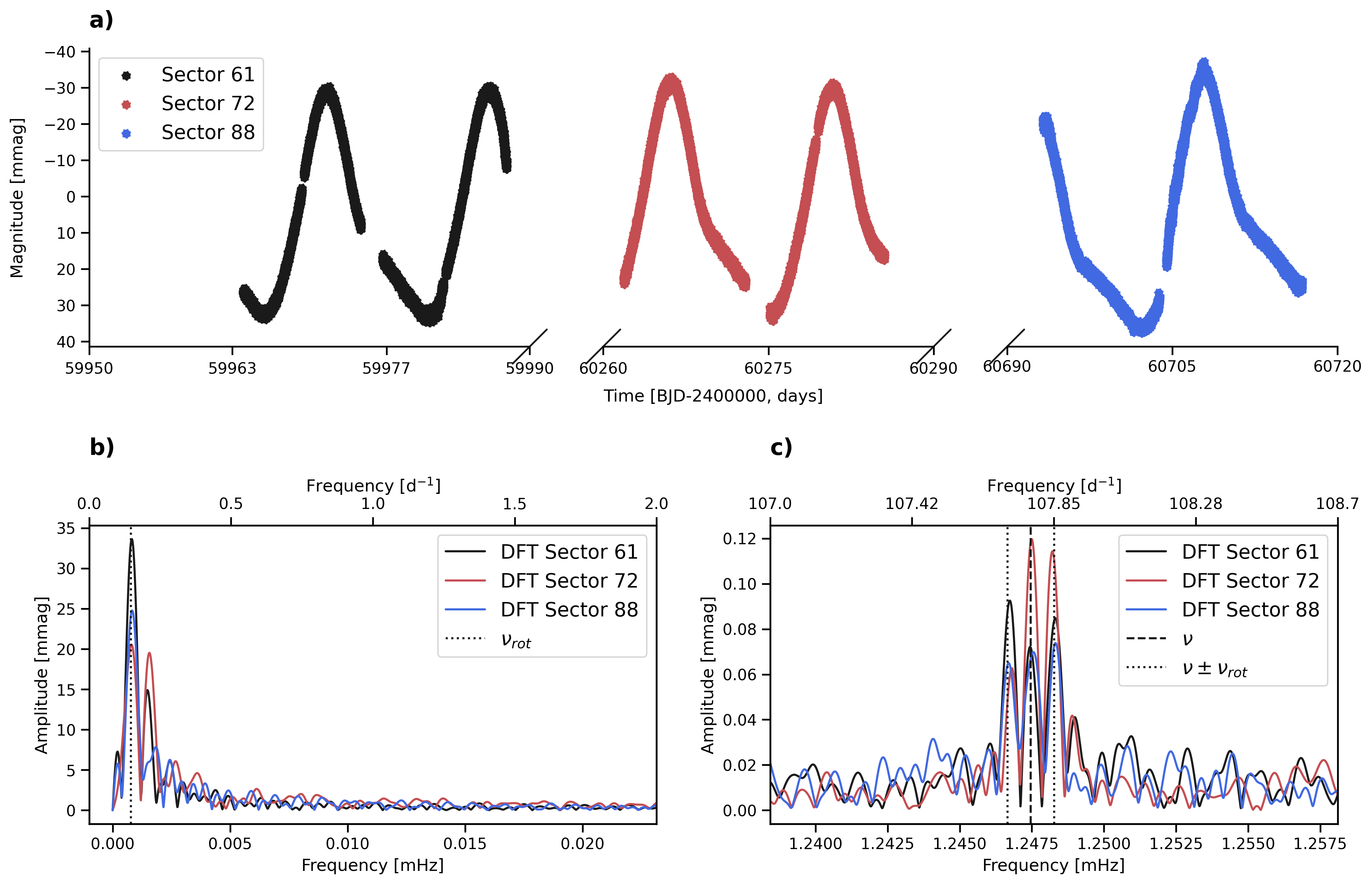}
    \caption{Results of the Fourier analysis performed on QLP data for TIC\,312111544. \textit{Panel a):} QLP light curves from TESS Sectors 61, 72, and 88. \textit{Panels b) and c)} show the DFTs of each sector, as well as the mean rotation (b, black dotted line) and the mean mode frequency obtained (c, black dashed line). The black dotted lines in \textit{panel c)} represent sidelobe frequencies. The frequencies highlighted in \textit{panel c)} are obtained after pre-whitening the rotation signals shown in \textit{panel b)}.}
    \label{fig:TIC312111544}
\end{figure*}

\begin{figure}[h!]
    \centering
    \includegraphics[width=\columnwidth]{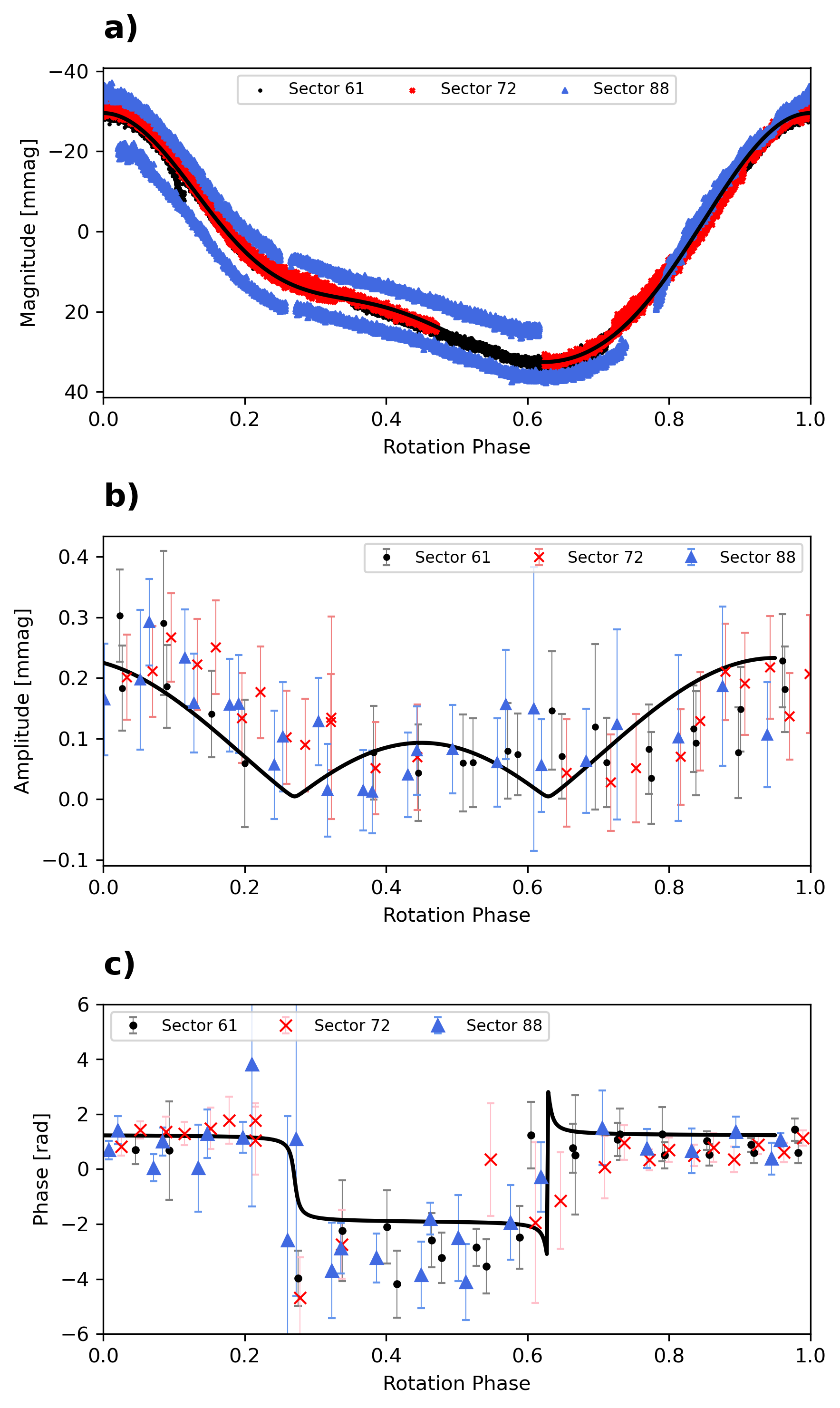}
    \caption{\textit{Panel a)}: Phase folded light curve of TIC 312111544 as observed in Sectors 61, 72 and 88 showing one rotational period ($P_{\rm rot}=14.7045$\,d). Phase zero corresponds to rotational light maximum at $59971.5158$ (BJD$-2400000$). \textit{Panels b) and c):} Amplitude and phase variability of the pulsation mode $\nu$ found by breaking the Sector 61, 72 and 88 light curves of into segments. The black solid line represents the theoretical amplitude and phase modulation for a pure dipole mode modelled according to \citet{1992Kurtz} using data from Sector 61. }
    \label{fig:phampvar}
\end{figure}

\begin{figure*}[h!]
    \centering
    \includegraphics[width=\textwidth]{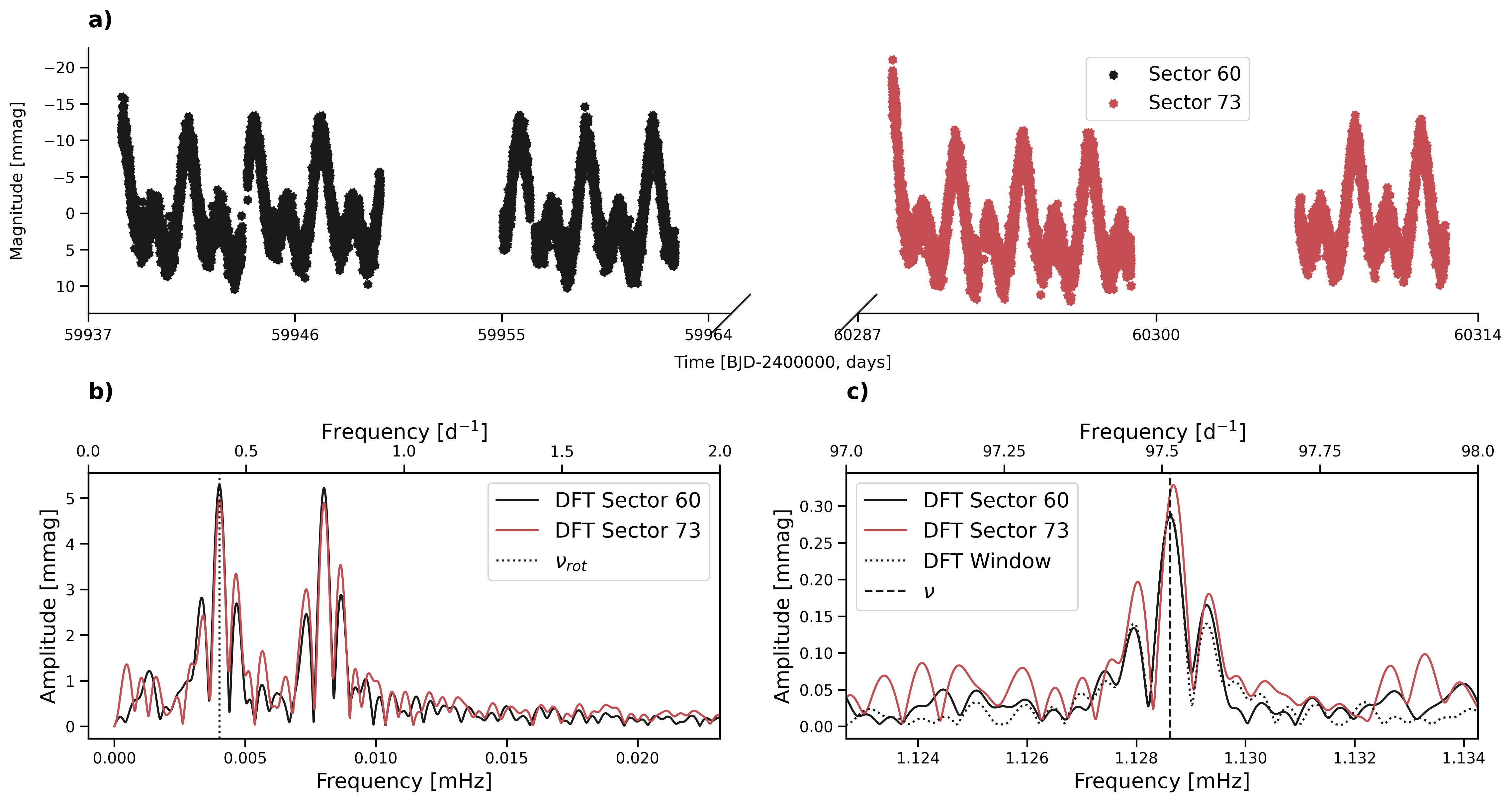}
    \caption{As Figure \ref{fig:TIC312111544} but for TIC 252881095. The black dotted line in panel c) enables visualisation of the window function of the DFT from Sector 60.}
    \label{fig:TIC252881095}
\end{figure*}

\begin{figure}[h!]
    \centering
    \includegraphics[scale=0.47]{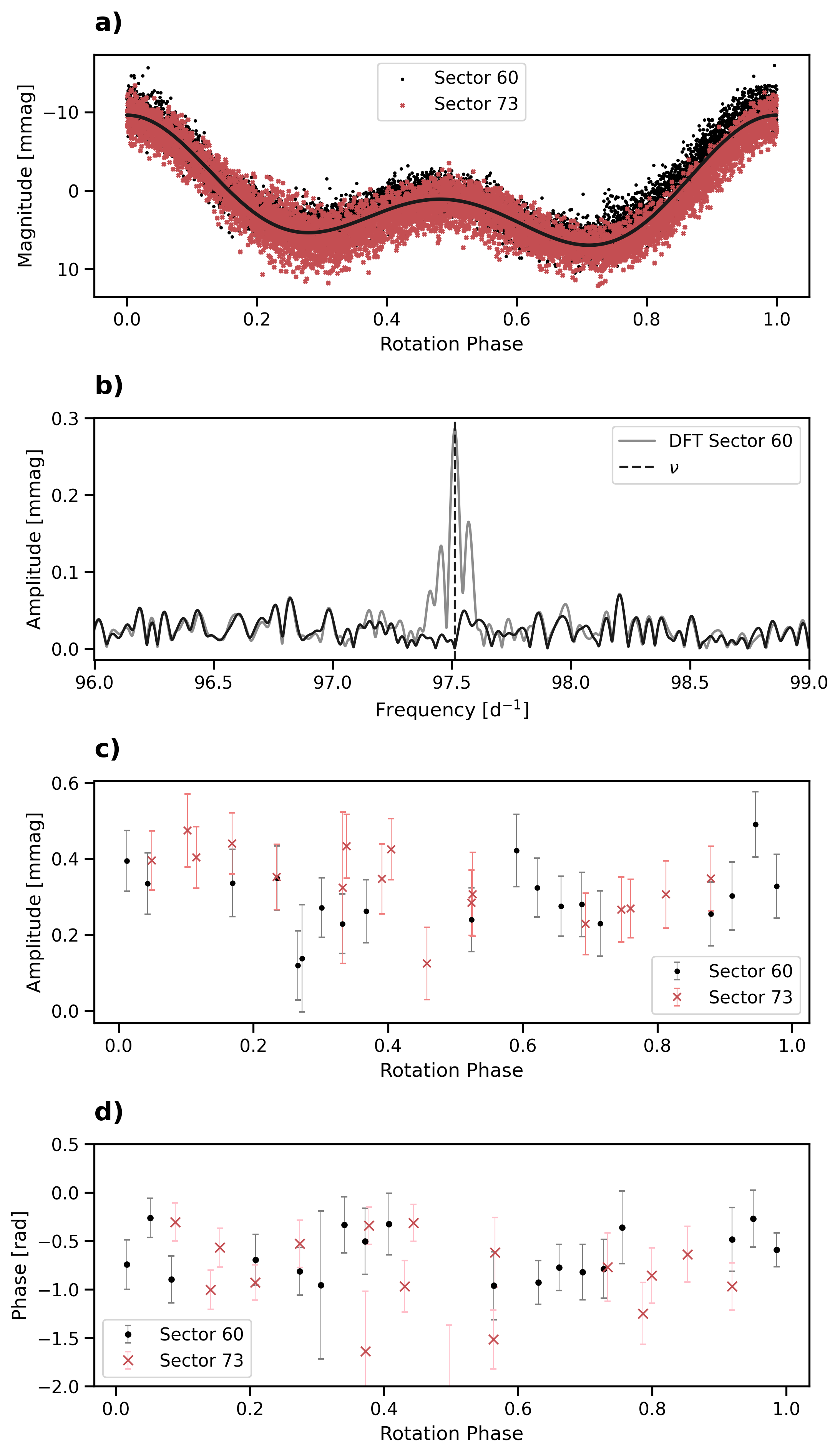}
    \caption{As in Figure~\ref{fig:phampvar}, but for the mode frequency $\nu$ of TIC\,252881095 ($P_{\rm rot}=2.8917$\,d). \textit{Panel a):} Phase zero corresponds to rotational light maximum at 59938.4869 (BJD$-2400000$). \textit{Panel b)} shows the DFT of TIC\,252881095 before (grey) and after (black) subtracting a least-squares sinusoidal fit at the mode frequency from the light curve.}
    \label{fig:phampvar_25}
\end{figure}

\begin{figure}[h!]
    \centering
    \includegraphics[width=\columnwidth]{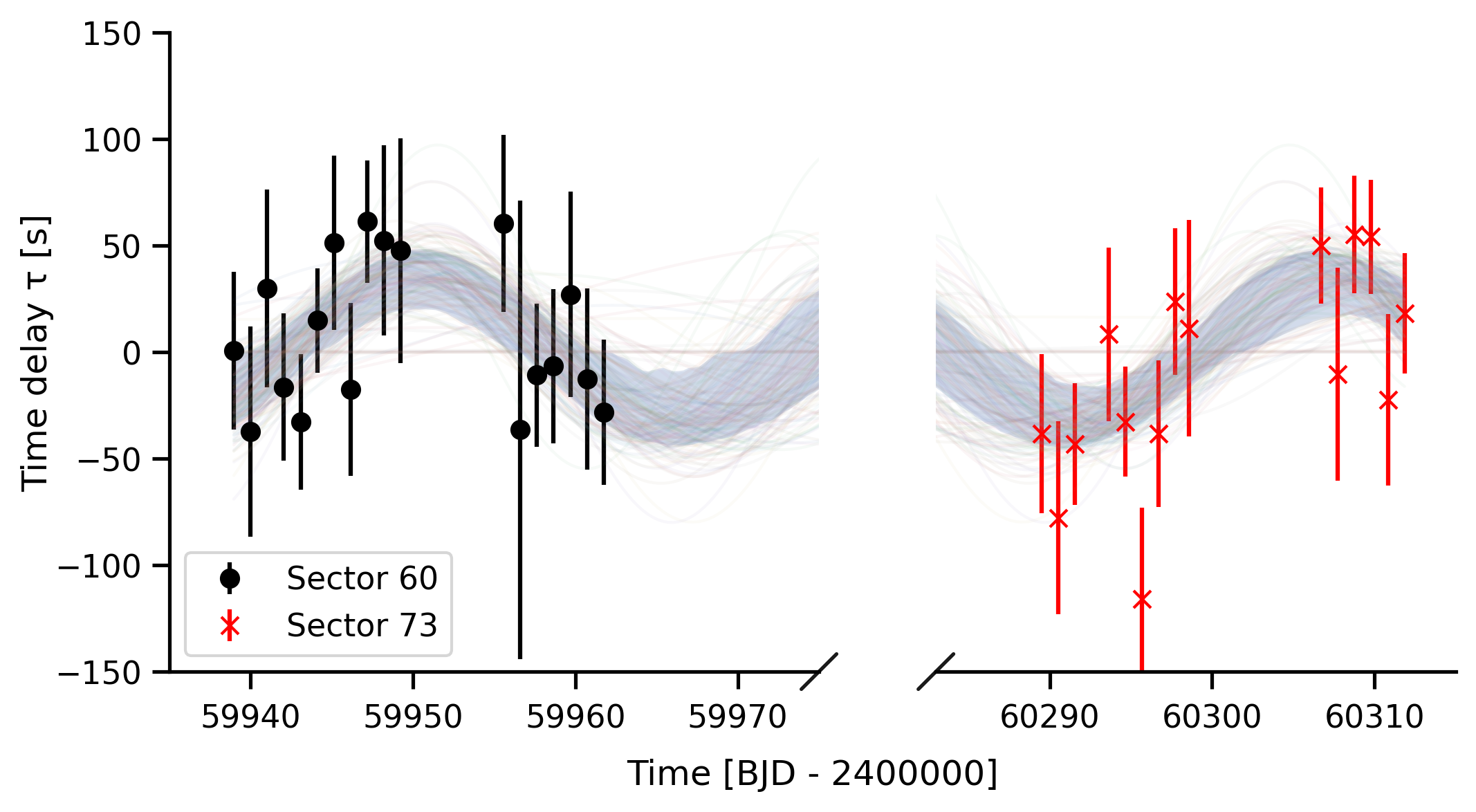}
    \caption{Most likely solutions of Equation~\ref{binary_eq} to fit the $\tau$ variability over time observed for TIC\,252881095.}
    \label{fig:tic25mcmc}
\end{figure}

\begin{figure*}[h!]
    \centering
    \includegraphics[width=\textwidth]{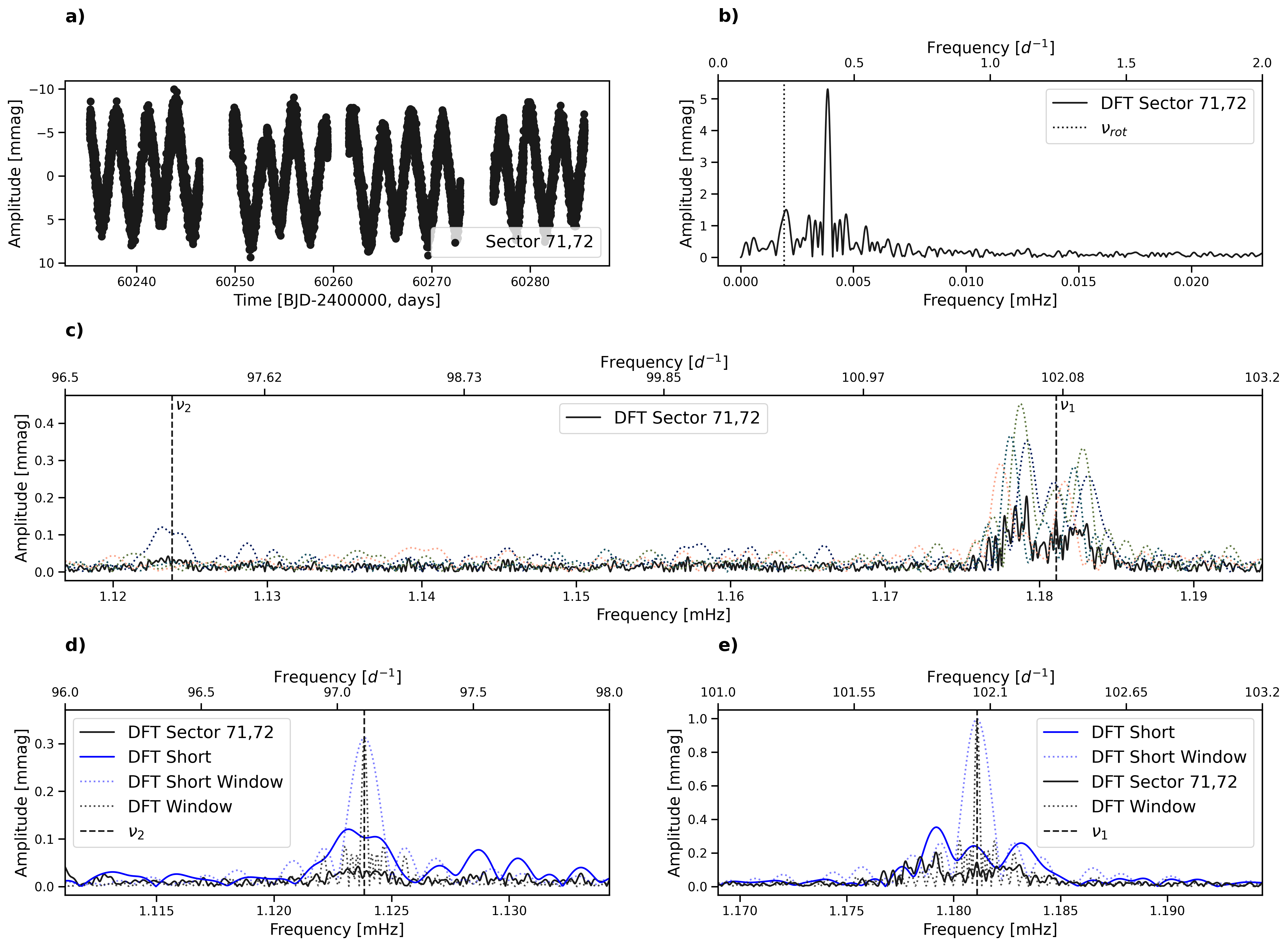}
    \caption{\textit{Panels a), b) and c):} As in Figure \ref{fig:TIC312111544} but for TIC\,46054683. \textit{Panel c)} additionally shows the DFT of each segment of the light curves (coloured dotted lines) used to study frequency/amplitude variability of the pulsation modes of this star. \textit{Panels d) and e)} show a zoom into the DFT regions around the two pulsation modes. The DFT computed from Sector 71 and 72 are shown in black. The blue line shows a DFT from a shorter segment spanning two rotation periods revealing two rotationally split sidelobes around the mode frequency. The dotted black lines enable visualisation of the window function of the DFT for the two mode frequencies.}
    \label{fig:TIC46054683}
\end{figure*}

\begin{figure}
    \centering
    \includegraphics[width=\columnwidth]{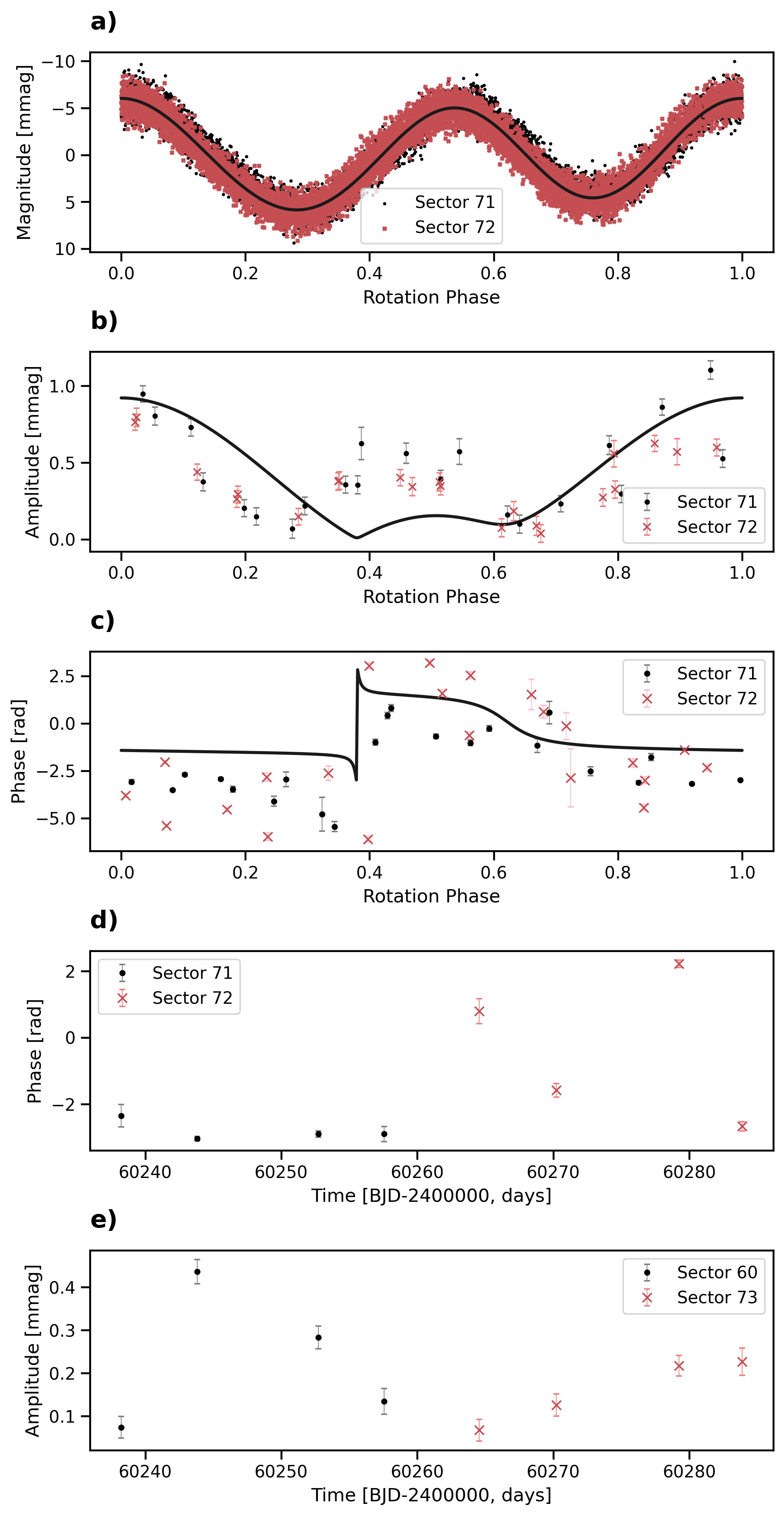}
    \caption{As in Figure \ref{fig:phampvar} but for $\nu_1$ of TIC\,46054683. \textit{Panel a):} Phase zero corresponds to rotational light maximum at 60237.8660 (BJD$-2400000$). \textit{Panels d) and e)} show the phase and amplitude variability over time, respectively.}
    \label{fig:46054683_phamvar}
\end{figure}

\begin{figure*}[h!]
    \centering
    \includegraphics[width=\textwidth]{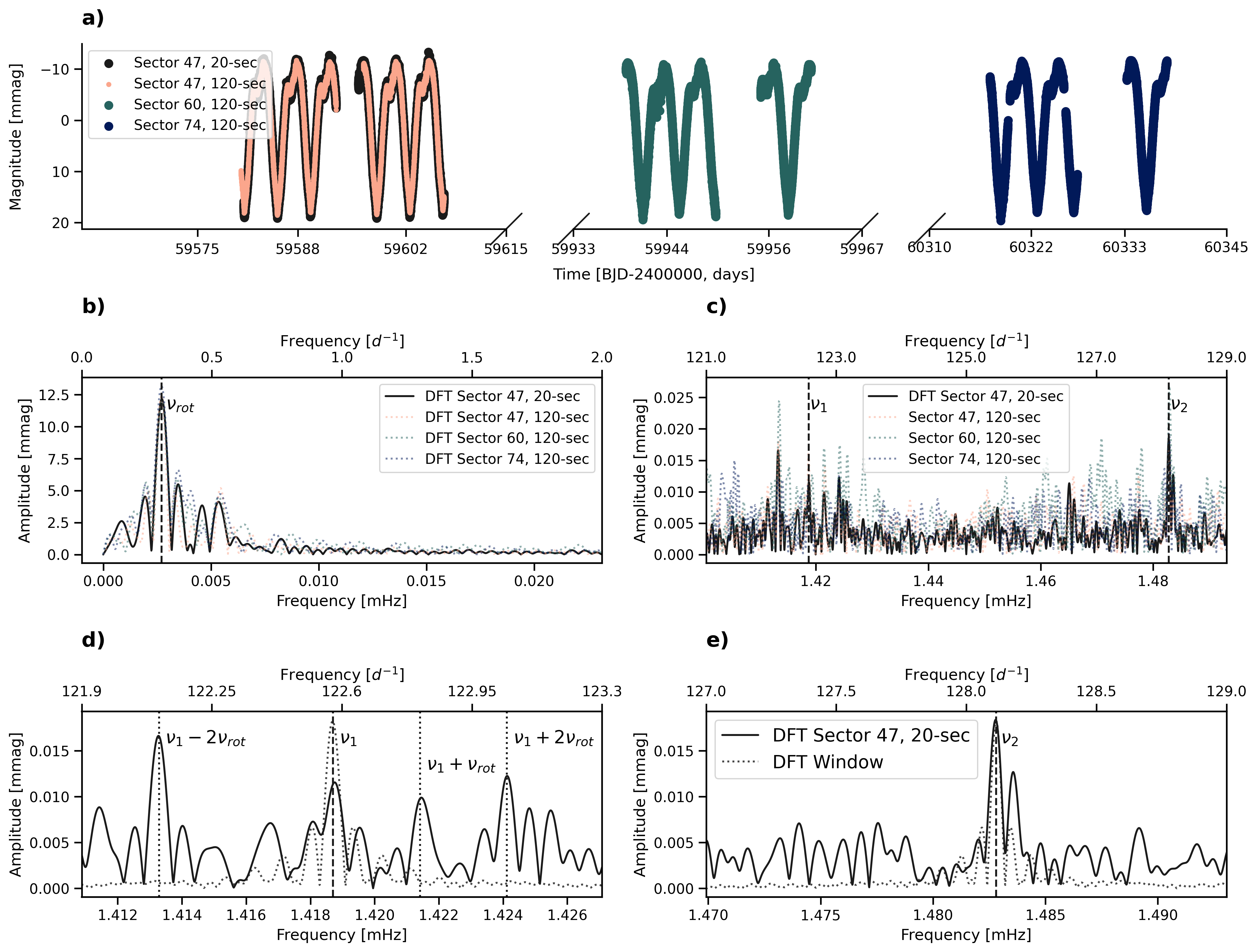}
    \caption{As in Figure \ref{fig:TIC46054683} but for 49 Cam (TIC\,393276640). Here, the coloured dotted lines show the DFTs calculated using the 120-s cadence data of the different Sectors.}
    \label{fig:TIC393276640}
\end{figure*}

\begin{figure}[h!]
    \centering
    \includegraphics[scale=0.475]{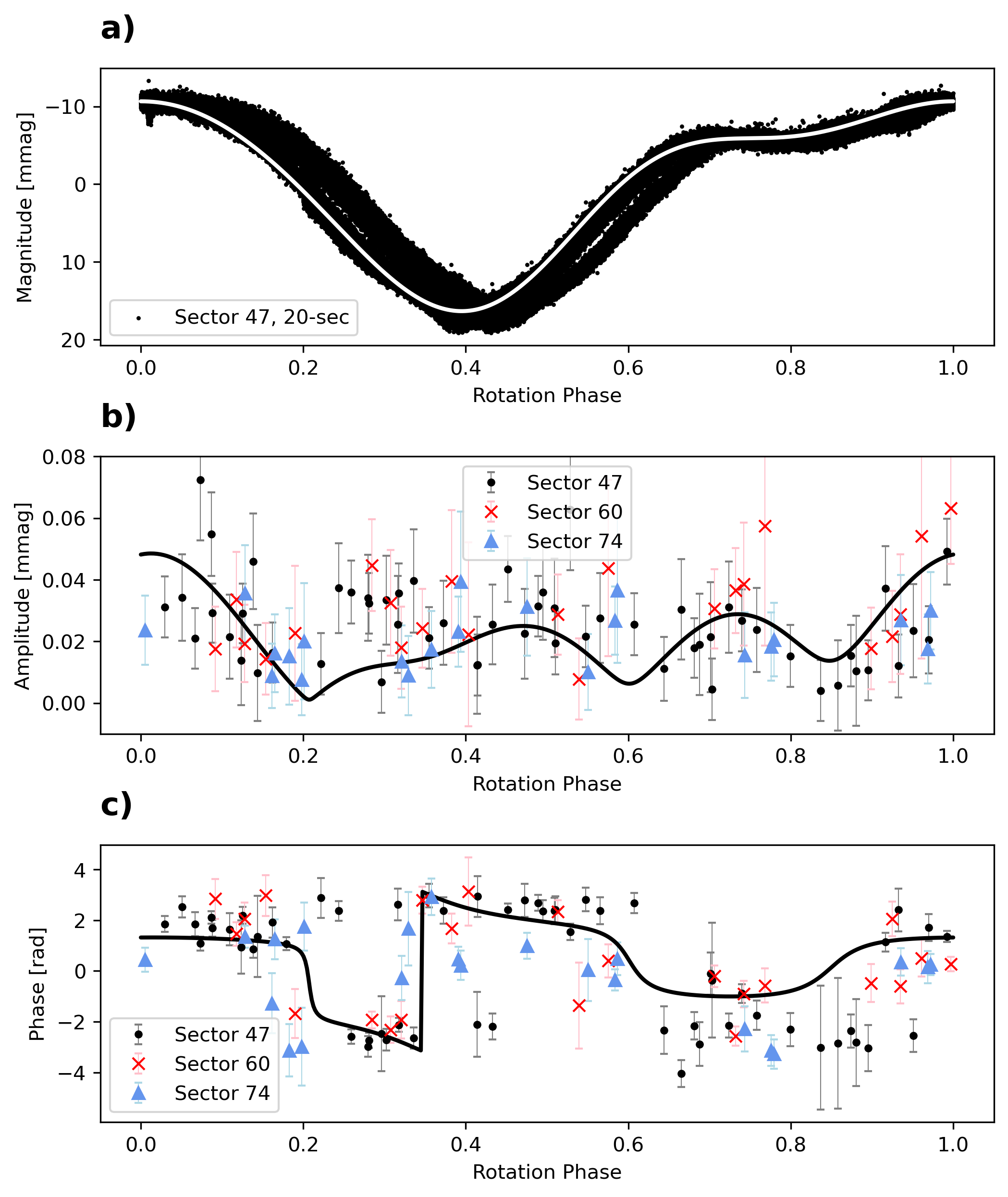}
    \caption{As in Figure \ref{fig:phampvar} but for $\nu_2$ of TIC\,393276640. \textit{Panel a):} Phase zero corresponds to rotational light maximum at 59583.8055 (BJD$-2400000$).}
    \label{fig:phampvar393276640}
\end{figure}

This method was developed as an alternative to the commonly used root-mean-squared (rms) noise estimate computed in the time domain. In the particular case of QLP data we retrieve the raw SAP flux for analysis, so light curves may still contain long-term trends, instrumental systematics and other artefacts. As such, for large-scale searches we estimate the noise level directly in the amplitude spectrum by identifying peak-free regions. A similar method was used in \citet{2024OliverandVichi} to identify variability in Am and Fm stars. Using QLP data for known roAp stars with well-established pulsation frequencies, we calibrated our frequency-domain noise estimation against the rms-based noise estimate and found an approximately one-to-one relation between the two. Our DFT-based noise estimate ($\sigma_{\text{DFT}}$) can therefore be interpreted in the same way as the standard rms amplitude noise, while being more robust against systematics found in QLP data. All frequencies with an amplitude measured to have SNR>4.5 $\sigma_{\rm DFT}$ are considered significant. Further details on the optimisation and performance of this approach are provided in Appendix~\ref{AOP}. 

The goal of this pipeline is to detect the high-frequency pulsations characteristic of roAp stars. Once new roAp candidates are identified by the pipeline, we return to the original light curves for each candidate and carry out a more detailed analysis, including the low-frequency regime. Our algorithm is thus set to classify stars as either non-pulsating (NP), roAp or $\delta$\,Scuti according to their pulsation frequencies. To that end we take advantage of the fact that different classes of variable stars show pulsations in different frequency ranges: $\delta$\,Scuti stars pulsate with frequencies of 4--80\,d$^{-1}$ \citep[e.g.][]{2004Amado,2014Antoci,chaplin2017asteroseismic,2019Antoci,2020Forteza,2024OliverandVichi}, while roAp stars pulsate with frequencies of 60--310\,d$^{-1}$ \citep[e.g.][]{2019Cunha,2020Mathys,2022Mathys,Cycle_1_2021,Cycle_2_2024}. Stars showing no significant frequencies in their amplitude spectra are classified as NP. 

Significant frequencies are identified in the DFT iteratively from highest to lowest amplitude and saved. The process stops when all peaks with amplitude above the required significance (SNR$>4.5\sigma_{\text{DFT}}$, Figure~\ref{fig:noise_win}, Appendix~\ref{AOP}) are found. Variable stars showing 100\% of the significant frequencies above 60\,d$^{-1}$ are automatically classified as roAp, while those showing 100\% of the significant frequencies below 60\,d$^{-1}$ are automatically classified as $\delta$\,Scuti. For stars with significant frequencies above and below 60\,d$^{-1}$, we test for combination frequencies. We restrict this to simple, first-order sum combinations: for each star we take the highest-frequency peak and check whether it can be reproduced by the sum of two other detected frequencies. When combination frequencies are found, the star is classified as $\delta$\,Scuti; otherwise as roAp.

\section{TESS 200-s cadence data of LAMOST Ap stars}
\label{200-s }

We cross-matched the $\sim$2700 stars classified as Ap in the LAMOST spectroscopic survey \citep[data releases (DR) 4-9; ][]{2020AandA...640A..40H,2020Shi} with the \tess\ target list. Among these, 1464 stars were observed with 200-s cadence. We applied our algorithm to this subset, and identified three new roAp stars: TIC\,312111544, TIC\,252881095, and TIC\,46054683.

\subsection{TIC 312111544}
\label{tic31}

TIC\,312111544 was observed in \tess\ Sectors 61, 72 and 88. We analyse the QLP light curves obtained in each Sector individually to derive relevant physical information regarding this star and how it pulsates (Figure \ref{fig:TIC312111544}). The frequencies ($\nu$), amplitudes ($A$), phases ($\phi$), and their respective errors were derived from an iterative multi-sinusoid non-linear least-squares fit to the time-domain light curves. The identified frequencies were adopted and successively pre-whitened until the residuals were consistent with white noise. Table~\ref{tab:obsall} shows the results obtained.

The QLP data show a clear rotational signal (panel a of Figure \ref{fig:TIC312111544}), allowing for the estimation of the stellar rotation period, $P_{\text{rot}}=14.7045 \pm 0.0004$ d, by combining the light curves of Sectors 61, 72 and 88 and fitting the low frequency range of the combined DFT. The International Variable Star Index (VSX) catalogue of the American Association of Variable Star Observers \citep[AAVSO, ][]{2006Watson} reports a period of $P_{\text{rot}}=14.724$\,d for this star, close to the one found in this work.  

Because the light curves from Sectors 61, 72 and 88 are separated in time, the window function introduces alias peaks ($\nu '$) near each true frequency in the DFT. To select the physical mode frequency ($\nu$), we choose the peak in the combined DFT that, unlike its aliases, minimises the phase variability over time. For each $\nu'$ we split the light curves into intervals of one rotation period, and fit each segment using least-squares to derive amplitude, phase and the respective errors. Breaking the light curve into $P_{\text{rot}}$ segments mitigates phase variability introduced by stellar rotation by avoiding the $\pi$-rad phase jump that occurs when the line of sight crosses a pulsation node on the stellar surface. Segments with too few points yield large phase errors, so we exclude any points with error larger than $\pi$. Using this method we identify one pulsation mode with frequency $\nu= 107.778 \pm 0.001$ d$^{-1}$, as well as 2 rotationally split sidelobes (panel b of Figure \ref{fig:TIC312111544}), in accordance with the oblique pulsator model.

If the observed multiplet is produced by rotational modulation of a single, non-distorted mode, i.e. well represented by a single spherical harmonic, then the multiplet components have fixed phase relations. For the time of pulsation maximum ($t_0$), we expect all components to be in phase. In this particular case, the splitting also determines the degree ($l$) of a given mode as multiplets are expected to have 2$l$+1 components. To probe whether or not this is a non-distorted mode, we assume the frequencies of the sidelobes to be separated from $\nu$ by exactly the rotation frequency ($\nu_{\text{rot}}$). We then force fit the sidelobes choosing $t_0$ appropriately by subtracting from the mean time, $\bar t$, a time offset $\delta t_{\pm1}$,

\begin{equation}
    \delta t_{\pm1}=\frac{\phi_m - \phi_n}{2\pi(\nu_m-\nu_n)}=\frac{\phi_{+1} - \phi_{-1}}{4\pi\nu_{\text{rot}}},
\end{equation}
where the subscripts $m$ and $n$ determine the phases/frequencies of the sidelobes \citep[Equation 26 in ][]{Kurtz1982}. Table \ref{tab:force_fit} shows the results of this process. The phases of the frequencies considered in all Sectors agree within $2\sigma_\phi$ errors, suggesting a non-distorted dipole ($l=1$) mode for TIC\,312111544.

For a non-distorted dipole mode we would also expect the amplitude of the mode frequency to drop to zero and its phase to shift by $\pi$-rad when the pulsation node crosses the line of sight \citep[e.g.][]{Kurtz1982,2016Holdsworth,2020Shi}. To infer amplitude and phase variability over the stellar rotational period, we follow a similar process to the one used to infer the true pulsation mode frequency of this star. However, since for this exercise we are interested in the variability rotation introduces in the pulsation phase, we split the light curve into shorter segments of 0.93 days ($\sim100/\nu$) to resolve it. The amplitude and phase variability of the pulsation mode $\nu$ over the rotation period are shown in Figure \ref{fig:phampvar}. The variability is clear in both cases, and the amplitude drops to zero as the phase shifts by $\pi$. These observations agree with theoretical amplitude and phase variability curves for a non-distorted dipole mode, computed from the Sector 61 results (Table \ref{tab:force_fit}) following \citet{1992Kurtz}. This supports our previous claim that this is a non-distorted dipole mode.

Observations suggest that TIC\,312111544 may be in a binary system. \textit{Gaia} reports a RUWE of $\sim 3.6$, well above the threshold of $>1.2$ often used to indicate unresolved companions \citep[e.g.][]{2020Berger2,2020Berger,2022Penoyre,2023GaiaCollab}. TIC\,312111544 is also listed in the WDS with a companion of $\Delta m \sim 4.45$, and the \textit{Gaia} DR3 catalogue shows two sources within 5 arcsec whose parallaxes differ by only $\sim$0.24 mas, suggesting they may form a physically bound pair (Figure~\ref{fig:TPF_312111544}). The primary has a Gaia magnitude of $G=9.23$ and was classified as kA3hA6mA9 SrCrEu by \citet{2020AandA...640A..40H}, indicating an Ap star with Ca {\sc{ii}} K, hydrogen, and metallic lines corresponding to spectral types A3, A6, and A9, respectively. Strong overabundances of Sr, Cr, and Eu are also evident in the LAMOST low-resolution spectrum (Figure~\ref{fig:lamost_blue}). The observed magnitude difference implies a much fainter secondary, consistent with a K-type main-sequence star. Additional data are required to confirm the binarity of this target. Such a possibility remains particularly interesting as roAp stars are predominantly single stars \citep[e.g.][]{2012AandA...545A..38S, Hartmann2015}.

\subsection{TIC 252881095}

\citet{2020AandA...640A..40H} classified this star as kA4hA6mF2 SrCrEu, indicating TIC\,252881095 is an Ap star with Ca {\sc{ii}} K, hydrogen, and metallic lines corresponding to spectral types A4, A6, and F2, respectively. Strong overabundances of Sr, Cr and Eu are evident in the LAMOST low-resolution spectrum (Figure \ref{fig:lamost_blue}). An unidentified line is also seen at $\sim 4379$\,\AA .

TIC\,252881095 was observed by TESS in Sectors 60 and 73 (Figure \ref{fig:TIC252881095}). The QLP data show clear rotational modulation, making it possible to measure the star's rotation period as $P_{\text{rot}}=2.8917 \pm 0.0001$\,d. The DFTs show another peak appearing at low frequency. This is the second harmonic of the rotation frequency, arising from the double-wave nature of the light curve. The double wave is likely caused by two chemical spots located on opposite poles of the stellar surface. As the \tess\ light curves are not consecutive, we combine them to determine the pulsation frequency following the method described in Section~\ref{tic31}. We also identify one pulsation mode at $\nu\sim97.513 \pm 0.001$ d$^{-1}$ (see Table \ref{tab:obsall}). The DFT shows additional peaks with a separation from $\nu$ not consistent with $\nu_{\text{rot}}$. These peaks can be attributed to the DFT window function, as illustrated by the synthetic signal (computed at the mode frequency for Sector 60) shown as the dotted black line in panel c of Figure \ref{fig:TIC252881095}. This is confirmed by the fact that when pre-whitening the light curve at the mode frequency these peaks disappear, leaving only noise (panel b of Figure \ref{fig:phampvar_25}). Although purely radial ($l=0$) modes are uncommon in roAp stars \citep[e.g.][]{2011Bigot,2019Holdsworth,2021Shi,2021FrASS...8...31H,Cycle_1_2021,Cycle_2_2024,2024Kurtz}, the current data show no evidence that this mode is part of a rotationally split multiplet, as the mode amplitude and phase show neither a minimum nor a $\pi$-rad phase shift associated with node crossing (Figure~\ref{fig:phampvar_25}).

TIC\,252881095 might be part of a binary system. This star is listed in both the WDS and the Tycho Double Star Catalogue \citep[TDSC,][]{2002Fabircius}. WDS flags it as a `Dubious Double', meaning the entry for TIC\,252881095 could be a positional typo or a spurious detection. By contrast, TDSC lists TIC\,252881095 as a resolved pair of two components. The Tycho photometry indicates that the two sources have nearly equal brightness ($\Delta m \sim 0.16$\,mag). However, at the \tess\ pixel scale such a pair would be fully blended, so the \tess\ light curve contains the combined flux of both components. The \textit{Gaia} DR3 RUWE value for this star is 1.20, equal to the threshold often used to flag unresolved binaries and consistent with the close pair reported in TDSC. 

We find two additional \textit{Gaia} sources close in projection to TIC\,252881095 (labelled 2 and 3 in Figure \ref{fig:TPF_252881095}). Source 2 (\textit{Gaia} DR3 991881778859792896) is only $\Delta m \sim$3 magnitudes fainter than the target, while source 3 is considerably fainter (\textit{Gaia} DR3 991881778858535040, $\Delta m \sim 9$). The \textit{Gaia} parallaxes indicate that the three stars are not at the same distance. TIC\,252881095 has $\varpi \simeq 1.40 \pm 0.02$\,mas, whereas source 2 has $\varpi \simeq 1.04 \pm 0.02$\,mas and source 3 has $\varpi \simeq 0.64 \pm 0.06$\,mas, both inconsistent with the target at $\gg 3\sigma$ when using the combined uncertainties. Thus, the nearby sources are most likely background objects. 

The rotation period of TIC\,252881095 is short enough to search for binary light-travel time effects in the \tess\ 200-s cadence data. Following \citet{2014Murphy}, it is possible to test whether a star is in a binary system by tracking the pulsation phase, $\phi(t)$, of a mode with frequency $\nu$ over time. For a binary system in a circular orbit, the corresponding time delay $\tau(t)$,
\begin{equation}
    \tau(t)=\frac{\Delta\phi(t)}{2\pi\nu},\label{tau}
\end{equation}
varies sinusoidally as
\begin{equation}
    \tau(t)=A_{\tau}\sin\left(\frac{2\pi t}{P_{\rm orb}} + \varphi\right) + C, \label{binary_eq}
\end{equation}
where $A_{\tau}$ is the maximum time delay, $t$ is time, $P_{\rm orb}$ is the orbital period, $\varphi$ is the orbital phase, and $C$ is a constant offset. 

We obtain $\Delta\phi(t)$ by dividing the light curve into segments spanning one rotation period, thereby averaging out the rotational phase dependence. For each segment, we perform a least-squares fit of a sinusoid to the time-domain data while fixing the frequency to the value $\nu$ measured from the full light curve. The resulting phases yield $\Delta\phi(t)$, from which we calculate $\tau(t)$ using Equation~\ref{tau}. We then fit Equation~\ref{binary_eq} to the resulting $\tau(t)$ measurements using nested sampling, exploring the parameter space under weakly informative priors (see Appendix~\ref{MCMC}). Figure~\ref{fig:tic25mcmc} shows the results for TIC\,252881095. The posterior is consistent with a sinusoidal solution for \(\tau(t)\) with an orbital period of \(P_{\mathrm{orb}}=32.41^{+7.84}_{-5.02}\)\,d (median and 68\% credible interval). Although most of the posterior mass concentrates on sinusoidal solutions, constant solutions cannot be ruled out. This is expected given the \(\tau(t)\) uncertainties and the gap between the two light curves.

To determine which model is preferred, we re-fit the data with two competing models: a sinusoid and a constant delay, both assuming the same noise model and similarly weak priors and compute the Bayes factor (BF). The evidence difference corresponds to BF$\sim$6.79, which, under equal model priors, implies the data favour the sinusoidal model with a posterior probability of $\sim 87\%$. In the Jeffreys/Kass–Raftery scale, this corresponds to moderate evidence supporting binarity. A longer continuous observation would be necessary to conclusively confirm TIC\,252881095 is part of a binary system. If confirmed, TIC\,252881095 would be among the few roAp stars known in close binaries \citep[e.g.][]{2012AandA...545A..38S, Hartmann2015}.

%\textcolor{blue}{DLH NOTES: Not much about this star in the literature. It is classified as Am by one LAMOST paper, and Ap in another. The spectrum looks like Ap to me, but there is a unknown line at $\sim 4379$\,\AA. It could be a blue shifted Fe line (4383\,\AA) but that is unlikely. It might be worth a comment. We think there is a multiplet in there; there is a lot of variability. How have you calculated the rotation period? Did you use both data sets? I derive a slightly different period.}

\subsection{TIC 46054683}
\label{tic46}

\citet{2020AandA...640A..40H} classified this star as A8 V SrCrEu, indicating TIC\,46054683 is a main-sequence Ap star of spectral type A8. The LAMOST low-resolution spectrum in Figure \ref{fig:lamost_blue} shows prominent features associated with Sr, Cr, and Eu, consistent with the reported overabundances of these elements.

TIC\,46054683 was observed in Sectors 71 and 72. As the two observations are consecutive, we combine the light curves for the analysis. The QLP data show a clear rotational signal (panel a of Figure \ref{fig:TIC46054683}), allowing us to measure a rotation period of $P_{\text{rot}}=6.0015\pm0.0009$\,d. This corresponds to the first low-frequency peak in the DFT (panel b of Figure \ref{fig:TIC46054683}). Another, more prominent peak appears at low frequency, corresponding to the second harmonic of the fundamental rotation frequency. This indicates that the rotational modulation is consistent with a double-wave light curve, likely caused by two chemical spots being located at opposite poles in the stellar surface.

The pulsations in this star are highly variable in both phase and amplitude. An amplitude spectrum of the combined and pre-whitened data is shown in panel c of Figure\,\ref{fig:TIC46054683}. There is clearly excess power around 100\,d$^{-1}$, but no coherent mode(s) identifiable. We therefore split the data into smaller sections to try and identify the pulsation mode(s). The results of this exercise are shown in panels d and e of Figure\,\ref{fig:TIC46054683}, where the light curve has been split into sections of two rotation periods and a DFT calculated for each. This reveals a three-peak structure around 100\,d$^{-1}$ that varies in both frequency, relative amplitude, and resolution. Determining a precise pulsation frequency for this mode is not possible given its variability, but we estimate the mode to be at $\sim 102$\,d$^{-1}$.

In an attempt to understand the variability further, we use this frequency estimate to track the pulsation amplitude and phase over the rotation period (panels a, b and c of Figure\,\ref{fig:46054683_phamvar}) and over time (panels d and e of Figure\,\ref{fig:46054683_phamvar}). We find that the variability over the rotation period is consistent with the theoretical modulation curves computed for a dipole mode \citep[$l$=1,][]{1992Kurtz}. Whether the mode is distorted is inconclusive from the available observational data. Furthermore, it is clear that the amplitude increases shortly after the observations start, but then begins to slowly decay over the 51.5 d of observations. The phase variations show a significant change over the observations, indicating that the frequency is highly variable in this star. This is not a new discovery in roAp stars \cite[see e.g.,][]{2021FrASS...8...31H}, but is perhaps the most dramatic case yet observed in which the pulsations do not vanish \cite[as in][]{2025MNRAS.536.2103K}. It is a combination of the amplitude and frequency (phase) changes that result in the incoherent power excess in the data shown in Figure\,\ref{fig:TIC46054683}.

There is evidence of a second mode in the data set that is only present in one of the amplitude spectra plotted (panel d of Figure\,\ref{fig:TIC46054683}). We estimate the mode frequency to be 97\,d$^{-1}$, leading to a separation between the two modes of $\sim 55\,\muup$Hz. This value is consistent with the estimated value of the large frequency separation for this star, $57\,\muup$Hz, calculated using the stellar parameters in the literature and scaling relative to the Sun \citep{Kj1995,Ulrich1986,EKER,2020AandA...640A..40H,2023GaiaCollab}. This supports the interpretation that the signal is a true mode rather than noise. Given both the relatively short time span of data, and the non-ideal cadence, it is imperative that TIC\,46054683 be observed in 20-s cadence mode for as long as possible in order to better characterise the nature of the transient pulsation modes in this star.

\section{TESS 20-s Cadence Data}
\label{20s}

Because roAp pulsations occur at high frequencies, short-cadence observations are ideal to avoid Nyquist-related ambiguity and to preserve sensitivity to additional low-amplitude modes and rotationally split sidelobes. Despite this, searches of roAp stars have mostly relied on longer-cadence data products, and the 20-s SPOC light curves remain relatively unexplored. To search for additional pulsation frequencies in known and candidate roAp stars, we adjusted the algorithm parameters described in Section~\ref{data} for the analysis of 20-s SPOC light curves. Namely, we increased the maximum frequency used in the computation of the DFTs to 2160~d$^{-1}$, which allows us to probe higher-frequency pulsation modes that may not be detectable in longer-cadence data. \textit{No detections were made above the 120-s Nyquist frequency (360 d$^{-1}$)}. However, the 20-s data revealed pulsations previously unobserved in the 120-s cadence data for one of the roAp candidates, allowing us to confirm its roAp star nature. The details of this new detection are given in the following section.

\subsection{49 Cam}

49 Cam (TIC 393276640) is a well-known Ap star, first classified as F0p (SrEu) by \citet{1968PCowley} and later as A8p (SrEu) by \citet{2000Leone}. The existence or lack of stellar pulsations in 49 Cam has long been a subject of interest \citep[e.g.,][]{2004Rya,2001Rya}. \citet{2001Rya} noted that roAp stars often show distinctive rare earth element (REE) abundance anomalies, where the abundances derived from the first and second ions of Pr and Nd differ by about two dex (Pr {\sc{ii}}–Pr {\sc{iii}} and Nd {\sc{ii}}–Nd {\sc{iii}}). Despite 49 Cam exhibiting this REE anomaly, former attempts to find oscillations from the ground have failed \citep[e.g.][]{1988Heller}. As such, 49 Cam was proposed for observation with TESS with both 120-s and 20-s cadence data.

49 Cam was observed in Sector 47 with a 20-s cadence and in Sectors 47, 60 and 74 with a 120-s cadence (Figure \ref{fig:TIC393276640}). The SPOC light curves show a clear rotation signal, from which we measure a rotation period of $P_{\text{rot}}=4.2375\pm0.0002$ d. The 120-s cadence data do not show any significant pulsation frequencies. Nevertheless, the analysis of the 20-s cadence light curve of 49 Cam revealed the existence of two high-frequency pulsation modes, establishing 49 Cam is a roAp star (Figure \ref{fig:TIC393276640}). From the DFT, we identify two pulsation modes at $\nu_1=122.576\pm0.003$ d$^{-1}$ and $\nu_2=128.115\pm0.002$ d$^{-1}$  (see Table \ref{tab:obsall}). The separation between the $\nu_1$ and $\nu_2$ modes is $\sim$\,64\,$\muup$Hz consistent with the theoretical value for the large separation frequency $\sim$\,53\,$\muup$Hz, estimated using the stellar parameters determined in \citet{2019Sikora_1} and scaling relative to the Sun \citep{Kj1995,Ulrich1986,EKER}.

The DFT shows two additional peaks adjacent to the mode frequency $\nu_2$ that are not separated from $\nu_2$ by the rotation frequency. These peaks remain present after pre-whitening $\nu_2$, at a significant level (SNR>4.5$\sigma_{\rm DFT}$, Figure~\ref{fig:TIC393276640}, panel~e). This residual structure is therefore likely related to amplitude and/or phase variability of the mode, as previously observed in roAp stars\citep[e.g.][]{Cycle_1_2021,Cycle_2_2024,2015MNRAS.452.3334S}.

Interestingly, $\nu_1$ appears to form a rotationally split quintuplet with a missing sidelobe. To test whether the multiplet is consistent with a non-distorted quadrupole mode, we perform a least-squares fit to the light curve, with the frequencies fixed to $\nu_1 + k\nu_{\rm rot}$ (with $k=-2,-1,0,+1,+2$; Table \ref{tab:force_fit}). In the standard case, the time of pulsation maximum, $t_0$, can be determined from the phases of the first sidelobe pair ($k=\pm1$), as these usually provide the strongest and most reliable phase constraint. However, for $\nu_1$ the $k=-1$ component is undetected in the DFT, so this pair cannot be used to define a meaningful $t_0$.
Instead, we determine $t_0$ from the detected outer sidelobe pair ($k=\pm2$) by subtracting from the mean time, $\bar t$, the time offset $\delta t_{\pm2}$,
\begin{equation}
    \delta t_{\pm2}=\frac{\phi_m - \phi_n}{2\pi(\nu_m-\nu_n)}
            =\frac{\phi_{+2}-\phi_{-2}}{8\pi\nu_{\text{rot}}},
\end{equation}
where $\phi_{+2}$ and $\phi_{-2}$ are the phases of the $\nu_1\pm2\nu_{\rm rot}$ components \citep[Equation 26 in ][]{Kurtz1982}. We then compare the phases of the remaining fitted components with that of the central frequency. After selecting $t_0$, the sidelobe phases remain inconsistent with the phase of the central component, indicating that $\nu_1$ is likely a distorted mode.

Following the methods of \citet{1992Kurtz} in Figure \ref{fig:phampvar393276640} we draw the theoretical curves expected for a quadrupole mode using the data from Table \ref{tab:force_fit}. Some scatter between different TESS Sectors is present across measurements obtained from different Sectors but the overall modulation pattern seems consistent with the theoretical curves. This supports that this star shows a quadrupole pulsation mode. Moreover, the amplitude modulation shows two maxima and two minima per rotation cycle consistent with the abrupt phase changes occurring near the amplitude minima as expected of a quadrupole pulsation in the oblique pulsator model. This is in agreement with the results of \citet{2017Silvester} that showed the magnetic field of this star is complex. In turn, such complex magnetic field could significantly contribute to the distortion of harmonic modes with $l$ up to 4. The observed scatter is best explained by the low amplitude of the mode, which makes the measurements noise-dominated when subdivided into shorter time segments. We also find no evidence for cyclical modulation of the amplitude or phase on timescales longer than the rotation period. Due to the low amplitude of this mode, and the fact that one rotationally split sidelobe of $\nu_1$ is missing we are not able to apply the oblique pulsator model to estimate $i$ and $\beta$ for this star.

This finding raises the question of whether the detectability of roAp pulsations depends on observational cadence.  \citet{2013PhDT.......638M} described this effect analytically using \textit{Kepler} data. For light curves of the same duration, shorter cadences provide more data points and therefore reduce the amplitude attenuation caused by time averaging over the finite integration time. This attenuation is frequency dependent, becoming increasingly important for higher-frequency pulsations, around the Nyquist limit. Thus, it is crucial to quantify how much information about the pulsation signal is lost when the data are sampled at longer cadences. We degrade 20-s SPOC light curves of two known roAp stars to effective cadences of 20–200 s and track the SNR of the dominant mode. For the faint TIC\,273777265 ($G=13.7$), the SNR decreases from 6.30 (20-s) to a value consistent with the degraded prediction in the original 120-s SPOC product, but drops to SNR$\approx$3 in the 200-s QLP data, below our detectability threshold. For the bright TIC\,280198016 ($G=6.2$), the SNR is lower in the original 120-s SPOC and 200-s QLP products than predicted from degradation of the 20-s data, most likely due to pipeline-dependent effects most pronounced for bright targets \citep{2022Huber}. Details of this exercise are in Appendix \ref{20secbin}. These results highlight the importance of short-cadence observations for reliable characterisation of pulsations in Ap stars.

\section{Conclusions}
\label{conc}

This work targeted 20-s and 200-s cadence \tess\ light curves to search for new roAp stars. We developed an algorithm to detect stellar pulsations in TESS light curves and applied it first to the 200\,s QLP data derived from TESS full-frame images (FFIs), focusing on a sample of Ap stars spectroscopically classified by LAMOST. This search revealed three previously unknown roAp stars: TIC 312111544, TIC 252881095, and TIC 46054683.

\citet{Cycle_2_2024} proposed that the incidence of roAp stars in the currently known population of Ap stars should be $\sim 5.5\%$. When exploring the LAMOST Ap sample, the detection rate in this work is inconsistent with the expected incidence. This discrepancy is likely related to the fact that our sample consists predominantly of faint targets, with an average \textit{Gaia} G mean magnitude of G=12.6 when compared with G=8.7 for the roAp stars reported in \citet{Cycle_1_2021,Cycle_2_2024} (see Appendix \ref{20secbin}). Furthermore, low-amplitude pulsations that could have been detected in shorter-cadence observations may be lost to noise in the 200-s cadence data. As shown in Appendix \ref{20secbin}, both the light curve cadence and the data processing pipeline significantly affect the SNR of detected pulsation modes. In general, 20-s SPOC light curves provide the highest SNR when compared with 120-s SPOC or 200-s QLP data, which are processed differently. In addition, when the 20-s cadence light curves are artificially degraded to lower cadences, their SNR decreases confirming that 20-s data yield the most reliable detection of high-frequency modes. This is supported by the discovery of two unreported pulsation modes in 49 Cam, thus confirming it is a roAp star.

We emphasize the importance of extended, consecutive temporal coverage for the characterisation of high-frequency modes in oblique pulsators (e.g., Sections \ref{tic31} and \ref{tic46}). Longer continuous observations increase the chance of resolving low-amplitude pulsation sidelobes and enable a more confident determination of the mode degree, $l$, which provides key insights into the stellar geometry and potentiates asteroseismic studies of these stars. In addition, extended coverage facilitates the detection of long-term cyclic frequency and amplitude variations, which can be indicative of binarity or other secular changes.

Overall, our results show that the combination of short-cadence TESS data, light curve processing, and temporal coverage impact the detection and characterisation of high-frequency pulsations in Ap stars. The discovery of new roAp stars, including previously unidentified modes in 49 Cam, are testaments to the potential of TESS to expand the known population of roAp stars and improve our understanding of their pulsation properties, mode geometries, and long-term variability. Continued monitoring of known and candidate roAp stars is also essential, as mode frequencies and amplitudes are well known to vary with time and, in the most dramatic cases, this variability can translate into intermittent detectability \citep[e.g.][]{2024Kurtz}. The origin of this variability remains poorly understood, and extended time-series photometry is therefore crucial to constrain its prevalence and timescales. 

Looking ahead, ESA’s PLAnetary Transits and Oscillations of stars mission \citep[PLATO;][]{Plato} will be particularly valuable for roAp studies, as its normal cameras will collect 25\,s cadence photometry and deliver light curves spanning $\sim$2\,yr. This combination of high cadence and long, continuous observations will enable more robust mode identification and long-term variability studies, while also improving sensitivity to low-amplitude pulsators. PLATO thus has the potential to substantially increase the number of detected roAp stars. Our automated search and classification pipeline is well-suited for application to PLATO data, paving the way for homogeneous, large-scale searches and characterisation of roAp stars in the near future.

\begin{acknowledgements}
      \scriptsize{IR, MSC and ARGS acknowledge support from Fundação para a Ciência e Tecnologia through the grants UID/04434/2025, 10.54499/2022.03993.PTDC and 2024.01287.BD, and work contracts with DOIs: 10.54499/2023.09303.CEECIND/CP2839/CT0003 and 10.54499/2020.02480.CEECIND/CP1631/CT0001. I.R. is also funded by Bolsas da Caixa Geral de Depósitos (CGD) by project the MIND PhD24 through the grant MP25PHD0136. ARGS also acknowledges the support from the GOLF and PLATO Centre National D'{\'{E}}tudes Spatiales grants. RJ is supported by the Klarman Fellowship. Co-funded by the European Union (ERC, MAGNIFY, Project 101126182 ). Views and opinions expressed are however those of the author(s) only and do not necessarily reflect those of the European Union or the European Research Council. Neither the European Union nor the granting authority can be held responsible for them. We thank the organizers of the 12th Iberian Meeting on Asteroseismology (IMAS) for fostering valuable interactions that contributed to this work. IMAS is an annual event organized by GIA (Group of Iberian Asteroseismologists; \url{https://gia.ugr.es/}). In particular, IR thanks Antonio García Hernández and Sebastiá Barceló Forteza for their helpful suggestions. We acknowledge the use of TESS High Level Science Products (HLSP) produced by QLP at the TESS Science Office at MIT, which are publicly available from MAST. Funding for the TESS mission is provided by NASA's Science Mission directorate. Guoshoujing Telescope (the Large Sky Area Multi-Object Fiber Spectroscopic Telescope LAMOST) is a National Major Scientific Project built by the Chinese Academy of Sciences. Funding for the project has been provided by the National Development and Reform Commission. LAMOST is operated and managed by the National Astronomical Observatories, Chinese Academy of Sciences. This work made use of scientific colour maps \citep[\url{https://www.fabiocrameri.ch/colourmaps/}]{crameri2020misuse} to mitigate visual distortion. }
\end{acknowledgements}

\bibliographystyle{aa}
\bibliography{Bibliography}

\clearpage

\begin{appendix}

This appendix contains supporting material that complements the main text.  

\section{Supporting material}
\label{A}

In this appendix we provide supporting material for the search algorithm described in Section~\ref{alg}, including its optimisation and performance tests, as well as additional light-curve and TPF \citep{2020TPF} diagnostics for selected targets.

\subsection{Search algorithm optimisation and performance}
\label{AOP}

\begin{figure}[h!]
    \centering
    \includegraphics[scale=0.5]{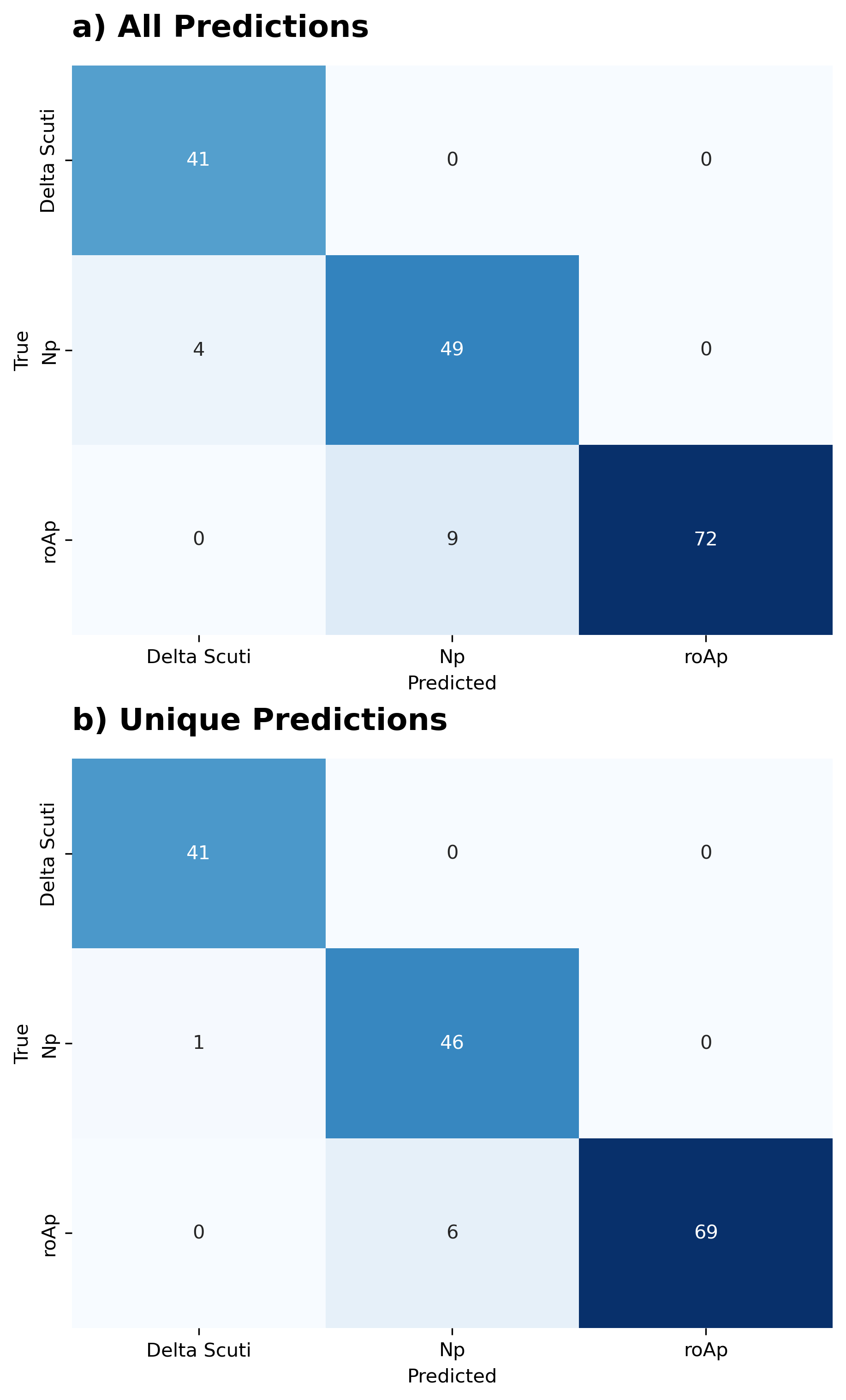}
    \caption{Confusion matrices summarising the performance of our search algorithm on a validation sample of 41 $\delta$ Scuti stars, 78 roAp stars and 50 non-pulsating stars. \textit{Panel a):} Results using all available predictions, including stars with different predictions for different sectors. \textit{Panel b):} Results removing stars with multiple classifications. Rows indicate the true classification and columns the predicted classification. The numbers in each cell give the classification counts.  }
    \label{fig:confusion}
\end{figure}

\begin{figure}[h!]
    \centering
    \includegraphics[scale=0.5]{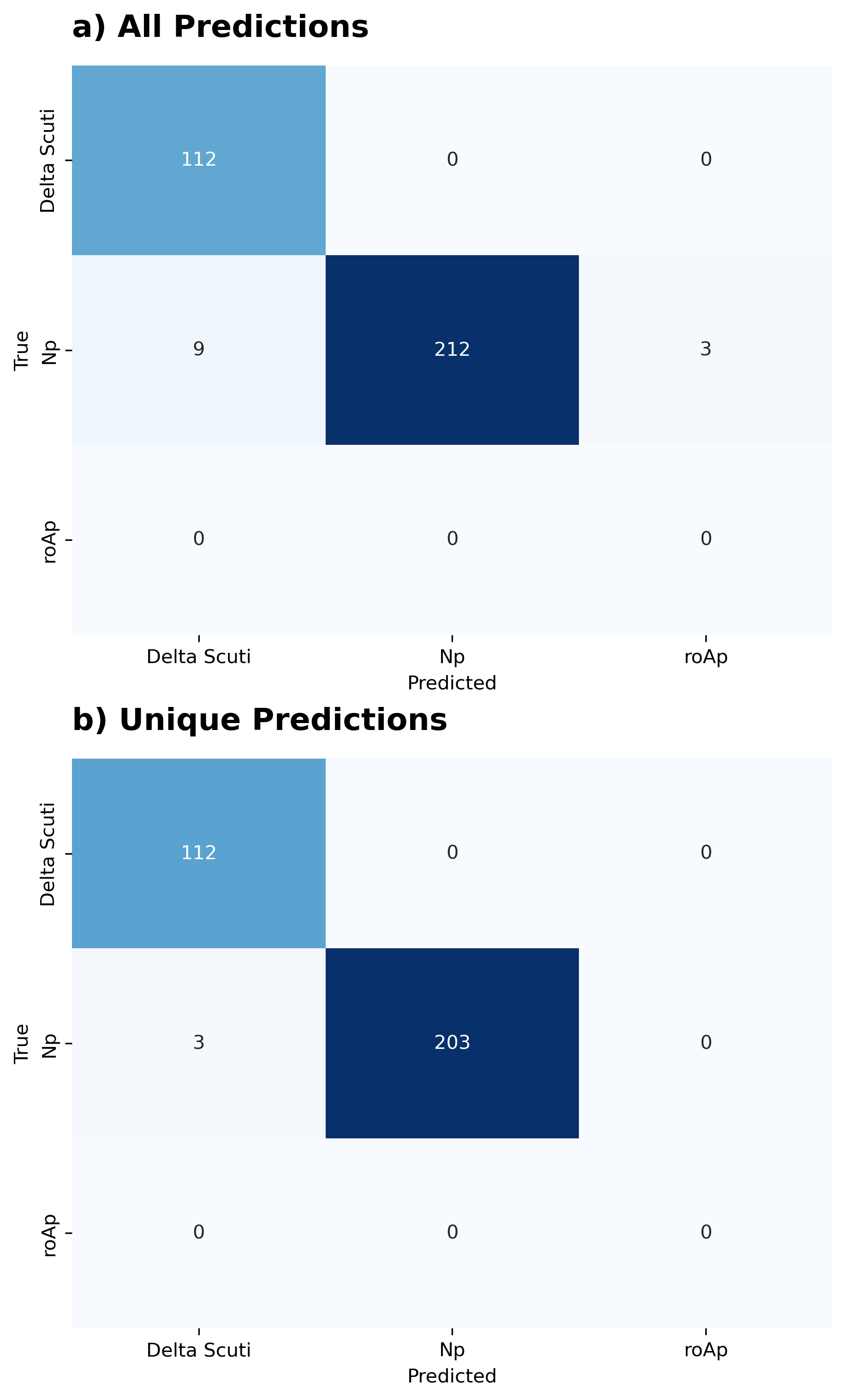}
    \caption{The same as Figure \ref{fig:confusion} on a test sample of 112 $\delta$ Scuti stars and 215 non-pulsating stars.}
    \label{fig:confusion_last_test}
\end{figure}

\begin{figure}[h!]
    \centering
    \includegraphics[scale=0.4]{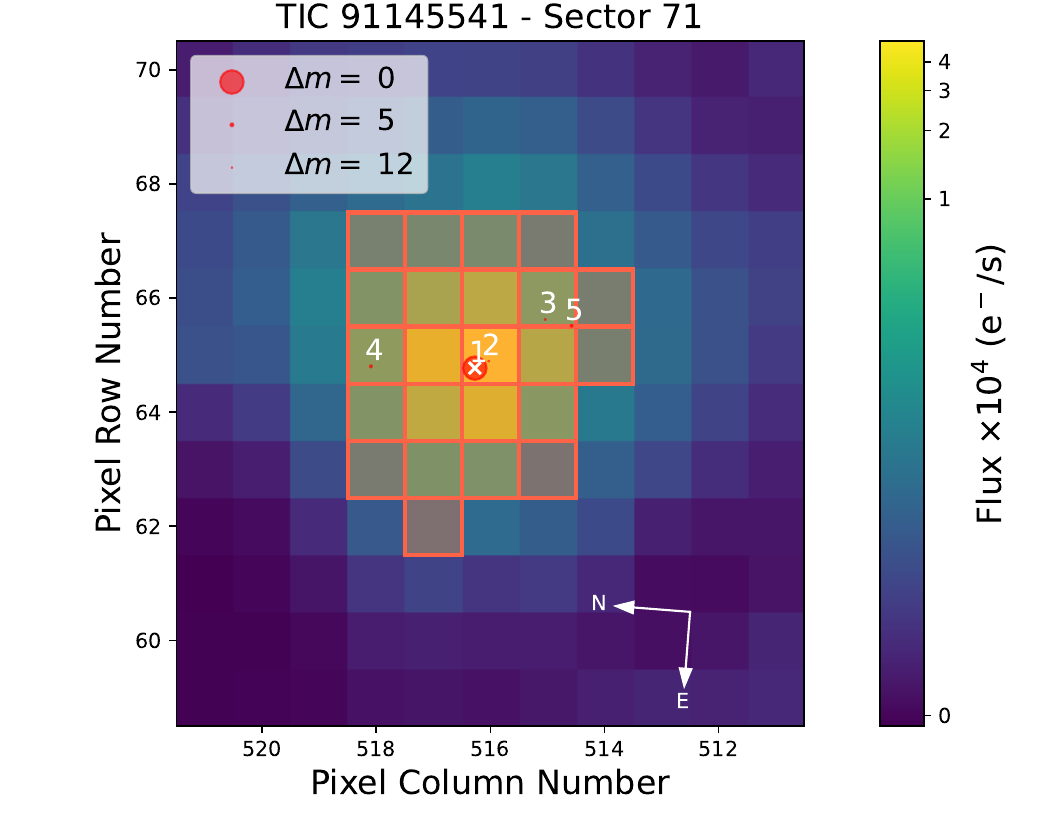}
    \caption{TESS target pixel file (TPF) for TIC\,91145541, observed in Sector 71. The background colormap shows the mean flux per pixel in units of 10$^3$ e$^-$ s$^{-1}$. The red outlines mark the chosen photometric aperture. The white cross indicates the target position ($\Delta m$ = 0), while nearby sources from \textit{Gaia} are shown as circles, sized according to magnitude difference. }
    \label{fig:binary}
\end{figure}

\begin{figure}[h!]
    \centering
    \includegraphics[width=\columnwidth]{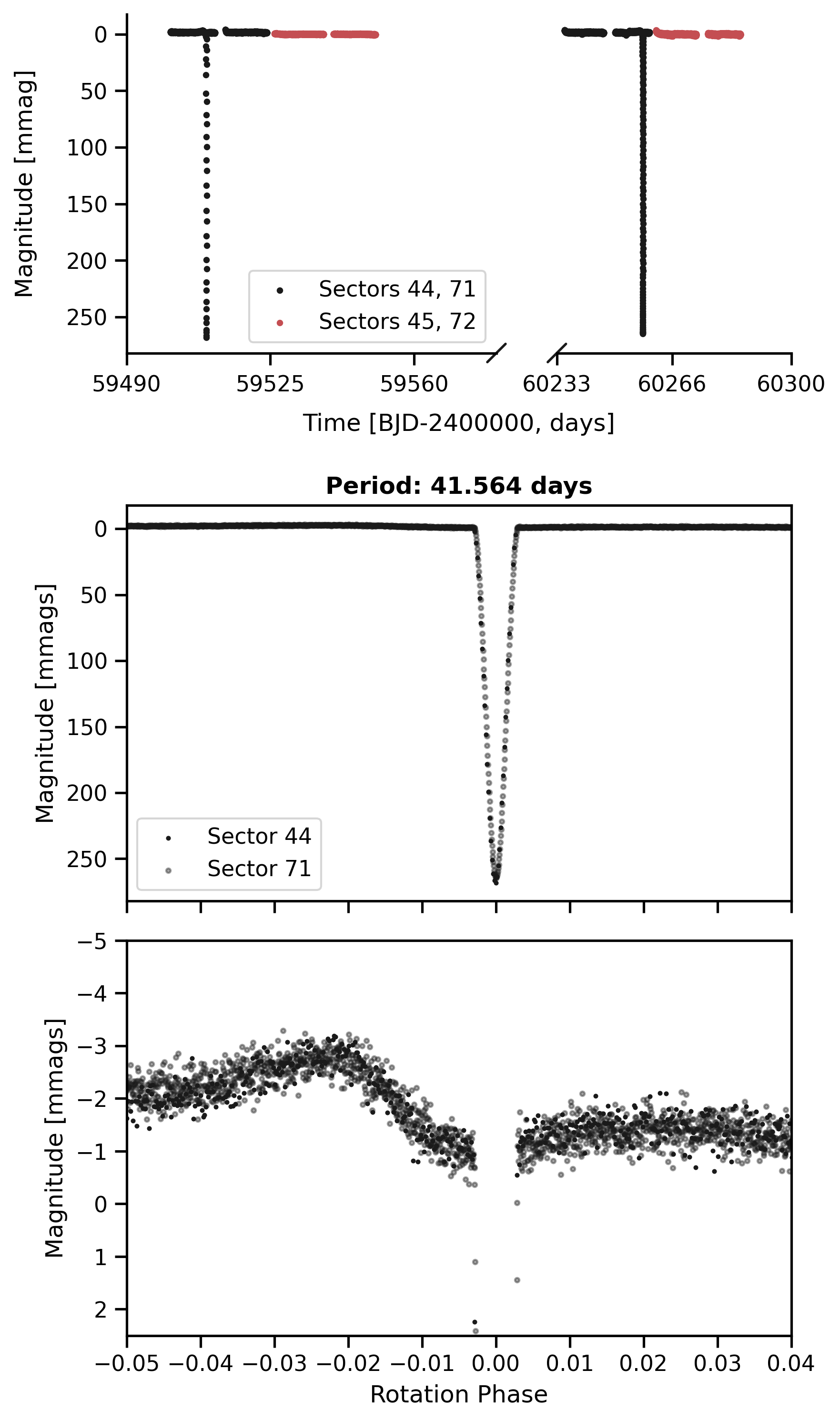}
    \caption{Eclipse observed in the QLP data for TIC\,91145541. The top panel shows the combined data of Sectors 44, 45, 71 and 72, providing a lower limit for the orbital period of the binary system of $\geq41.5$ days. In the \textit{middle} and \textit{bottom} panels are zoom-ins of different regions of the eclipse. The \textit{bottom} panel reveals an increase in brightness immediately before the eclipse, consistent across Sectors 44 and 71.}
    \label{fig:transit}
\end{figure}

%\clearpage

\begin{figure}[h!]
    \centering
    \includegraphics[scale=0.5]{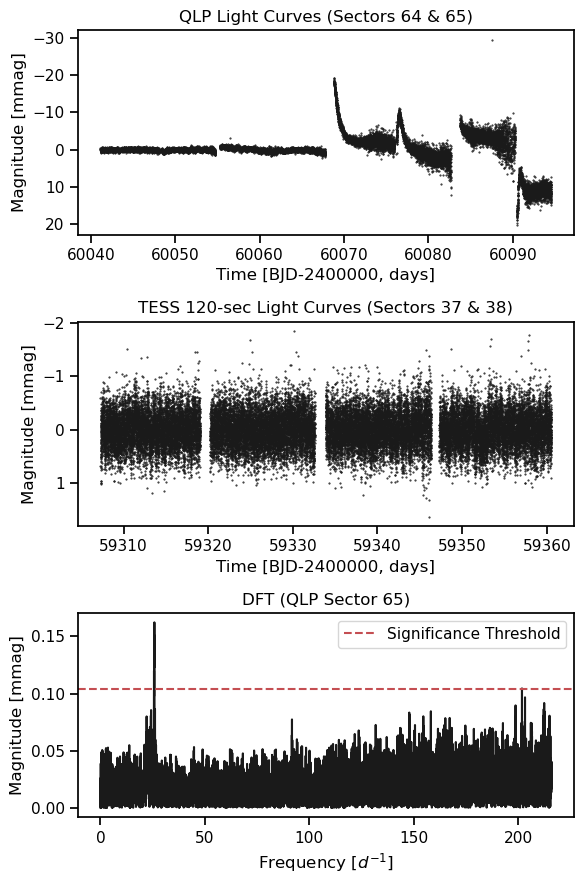}
    \caption{Variability observed for TIC\,381942100. \textit{Top panel:} Raw (SAP) flux from QLP light curves from TESS Sectors 64 and 65 (200-s cadence), showing the difference in variability between the highly variable light curve of Sector 65 compared to the nearly flat light curve of Sector 64. \textit{Middle panel:} Processed (PDC SAP) flux from SPOC 120-s cadence light curves from Sectors 37 and 38. \textit{Bottom panel:} Discrete Fourier Transform (DFT) of the QLP Sector 65 data, where several peaks are visible above the noise level. The dashed red line marks the adopted significance threshold.}
    \label{fig:TIC381942100}
\end{figure}

\begin{figure}[h!]
    \centering
    \includegraphics[scale=0.4]{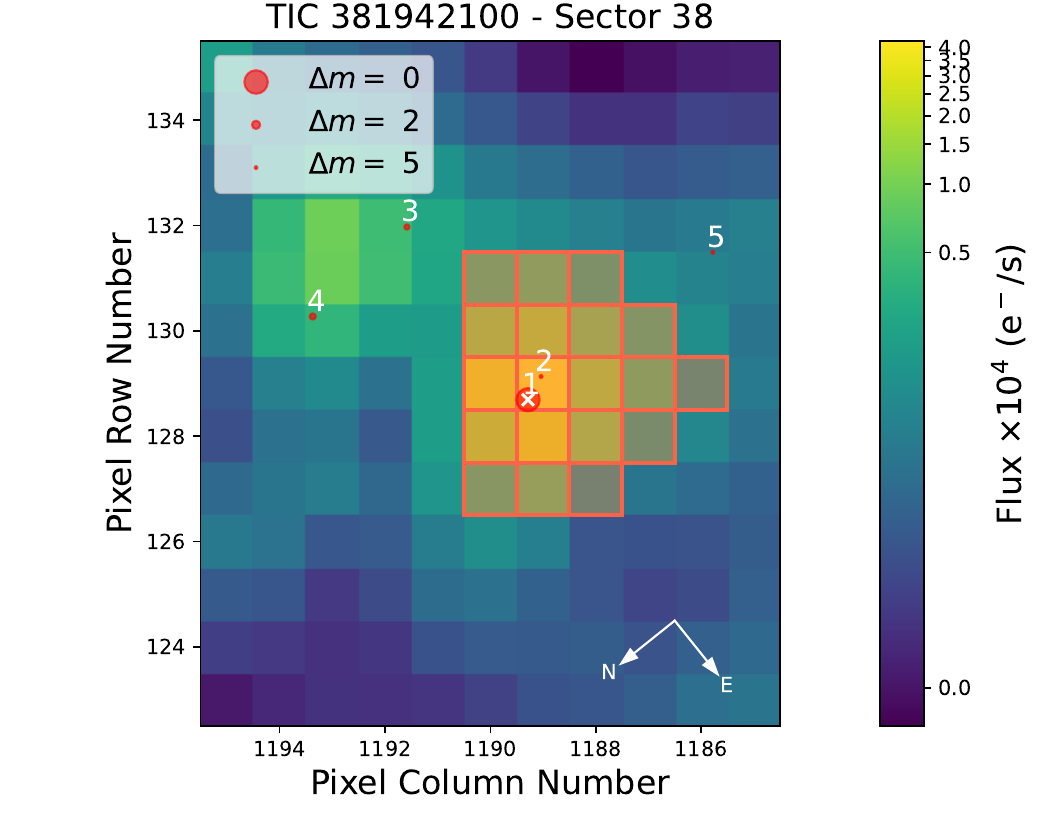}
    \caption{As Figure \ref{fig:binary} but for TIC\,381942100 observed in Sector 38.}
    \label{fig:TPF_381942100}
\end{figure}

To ensure our classification is robust against noise and false detections, we optimised the SNR threshold used to determine which frequencies in the DFT are considered significant. This was done by testing the algorithm on a sample of 120 stars with known variability types: 41 $\delta$ Scuti stars \citep{2019Antoci}, 78 roAp stars \citep{Cycle_1_2021,Cycle_2_2024}, and 50 non-variable stars \citep{2019Antoci,2020Mathys}. All stars in the testing sample have 120-s cadence data available, and all classifications obtained from the 120-s light curves are consistent with those reported in the literature.

These tests were performed using 200-s QLP cadence light curves, since these are the data products to which the search pipeline is applied. When discrepancies appear in the QLP classifications, we use the corresponding 120-s SPOC light curves, where available, as an external consistency check to help determine whether the detected signal is likely astrophysical or instead due to noise, systematics, or other artefacts. When analysing high-frequency roAp pulsations, QLP data generally have lower SNR than light curves processed with SPOC, such as the 20-s and 120-s cadence data. Additionally, the longer integration time of the 200-s data leads to an undersampling of the pulsation signal, reducing the observed amplitude. Nonetheless, the large quantity of QLP light curves makes them an important resource for large-scale roAp star searches.

We find that adopting a threshold of SNR$>4.5\sigma_{\rm DFT}$ reduces the number of false detections due to noise while preserving the detectability of most pulsation frequencies. At this threshold, all $\delta$\,Sct stars are correctly classified, 69 out of 78 roAp stars are classified as roAp, and 46 out of 50 NP stars are classified as NP (Figure~\ref{fig:confusion}). Of the 69 correctly classified roAp stars, 3 show sector-to-sector inconsistencies, being classified as roAp in some sectors and as NP in others. 

We additionally compared the performance of our adopted threshold with thresholds of 4.0 and 5.0. The lower threshold of 4.0 gives the highest roAp recovery rate (89.8\%), but also the highest false-positive rate among NP stars (6\%) and the largest fraction of sector-to-sector inconsistent classifications (20.1\%). The more conservative threshold of 5.0 reduces the roAp recovery rate to 84.1\%, while offering little improvement over the 4.5 threshold for NP stars. The adopted threshold of 4.5 therefore provides the best balance between completeness and robustness, with a roAp recovery rate of 87.5\%, a NP false-positive rate of 2\%, and an inconsistent sector-to-sector classification rate of 10.6\%. This is consistent with the conclusion of \citet{2021AcA....71..113B} that the commonly used SNR$=4$ is too permissive for space-based photometry, while showing that, for our 200-s QLP data and DFT-based local noise estimate, a threshold near 4.5 provides a better compromise than a more conservative threshold of 5.0.

After visual inspection of all light curves, we conclude that the algorithm performs according to expectation: all classifications match the analysed signal. In the case of the misclassified roAp stars, there are no significant peaks in the QLP amplitude spectrum that would lead to a roAp classification. This is expected to a certain extent, as the lower SNR of QLP light curves, compared with 120-s SPOC light curves (Appendix~\ref{20secbin}), can cause weak pulsation peaks to be lost in the noise. Since the SNR depends strongly on light-curve quality, it is also reasonable that low-SNR peaks may be detected in some sectors but not in others. Different classifications from different sectors may also indicate temporal evolution of the stellar pulsations themselves, as has been observed previously, for example in TIC\,340006157 \citep{2025MNRAS.536.2103K}. As for the misclassified NP stars, significant peaks in the $\delta$\,Scuti frequency range are found in some QLP light curves, but these are not recovered in the corresponding 120-s SPOC data and are therefore most likely instrumental or other non-astrophysical artefacts.

The optimisation exercise described above has shown that our algorithm is capable of detecting roAp pulsations from 200-s QLP data. However, because the optimisation sample is not representative of the $\delta$ Scuti and non-variable star populations, it is not suitable for evaluating the false-positive rate. To address this, we carried out an additional performance test on a sample of 112 $\delta$ Scuti stars classified by \citet{2024OliverandVichi}, and 215 non-variable stars, in order to assess the false-positive rate and ultimately the suitability of our algorithm for large-scale surveys using 200-s QLP light curves. We chose only stars in the roAp temperature range (6000--9000\,K). The sample of non-variable stars was built using 147 metallic-line A-type (Am) stars classified as non-variable by \citet{2024OliverandVichi}, together with 66 well-known non-variable Ap stars \citep[e.g.][]{2008Frey,2012Elkin,2019Chojnowski,2022Jarv,2022Roman,2023Rustem,2024Mathys,2025Mathys}.

In this test, all $\delta$ Scuti stars are correctly classified, 212 out of 215 NP stars are correctly classified as NP and 3 are misclassified as $\delta$ Scuti (Figure~\ref{fig:confusion_last_test}). Visual inspection of the corresponding signals suggests that these misidentifications are most likely noise artefacts. Of the 212 correctly classified NP stars, 9 received different classifications when data from different sectors were used. In most cases, these inconsistencies are likely due to sector-to-sector variations in the light-curve quality but on occasion these inconsistencies can be attributed to true physical signals. Two illustrative examples are discussed below.

\subsection{Illustrative cases}
\label{app:problem_cases}

TIC\,91145541 is an Am star of spectral classification kA3hA9mF2 \citep[IV,][]{2020McG} that was observed with 10-min cadence in Sectors 44 and 45 and 200-s cadence in Sectors 71 and 72. It shows a clear eclipse in the light curves of Sectors 44 and 71 (Figure~\ref{fig:transit}). The eclipse depth suggests $R_1/R_2\sim0.27$, implying that TIC\,91145541 is part of a binary system. The time between the two observed eclipses must be an integer multiple of the true orbital period. Given the observed separation of $\sim748$ days between eclipses, at least 18 full orbits must have occurred, placing a lower limit on the orbital period of $\geq41.5$ days. The star also shows an increase in brightness immediately before eclipse in Sectors 44 and 71, resembling the periastron brightening seen in heartbeat binaries. The eclipse signal introduces power at frequencies beyond the pre-whitening range (0--10\,d$^{-1}$), leading to the false $\delta$\,Scuti classification.

TIC\,91145541 had previously been flagged as a possible binary by \citet{2019Kervella} using \textit{Gaia} DR2 proper-motion anomalies. Using \texttt{tpfplotter} \citep{2020TPF}, we plotted the five nearest projected \textit{Gaia} sources over the \tess\ TPF for Sector 71 (Figure~\ref{fig:binary}), but do not resolve any clear companion. One nearby source in the \tess\ field of view (\textit{Gaia} DR3 3381547568847559808; object number 2 in Figure~\ref{fig:binary}) lies very close in projection to TIC\,91145541, but the parallax difference of 6.12 mas indicates that it is a background star rather than a physically bound companion.

For the remaining stars with inconsistent classifications, sector-to-sector variations in light-curve quality are the most likely explanation. In practice, when more than one light curve is available for a given target, we inspect whether the pulsation signal is consistently detected across different sectors. Signals that are not reproducible are treated as likely spurious detections. For the final interpretation we adopt the classification supported by the highest-quality available light curve. A clear example is TIC\,381942100 (Figure~\ref{fig:TIC381942100}), an Ap star of spectral type ApSrCr(Eu) \citep{1975mcts.book.....H} observed with 200-s cadence in Sectors 64 and 65. The QLP light curves differ dramatically: the Sector 65 data show pronounced flux variability, whereas the Sector 64 light curve appears much less variable. Accordingly, the star is classified as NP in Sector 64 but as roAp in Sector 65. The DFT of the Sector 65 data shows several significant frequencies in the $\delta$\,Scuti range and one peak just above the significance threshold in the roAp range. This apparent roAp detection is most likely caused by the rising high-frequency noise level in this light curve.

This star is listed in both the Catalogue of the Components of Double and Multiple Stars \citep[CCDM;][]{CCDM} and the Washington Double Star Catalog \citep[WDS;][]{2001WDS}, with a reported nearby source of magnitude difference $\Delta m\sim4.6$. However, we find no evidence that TIC\,381942100 is part of a physical binary system. The \textit{Gaia} RUWE is low (0.82), and the nearby sources identified in the TPF (Figure~\ref{fig:TPF_381942100}) all have parallaxes consistent with background objects. In particular, the source reported in CCDM and WDS is TIC\,381942077 (object 2 in Figure~\ref{fig:TPF_381942100}), whose \textit{Gaia} DR3 parallax differs by 7.65 mas from that of TIC\,381942100. Furthermore, \citet{2019Kervella} found no evidence for binarity from \textit{Gaia} DR2 proper-motion anomalies. We therefore retain the non-pulsating classification obtained from the Sector 64 QLP data and interpret the Sector 65 QLP roAp detection as spurious. The 120-s SPOC light curves from Sectors 37 and 38 show much better agreement in flux variability than the QLP data, and the DFT computed from those data supports a non-pulsating classification, in agreement with the Sector 64 QLP result.

%\clearpage

\onecolumn

\begin{landscape}
\section{Seismic Properties of new roAp stars}
\label{roap_tables}

In this section we present the derived seismic properties of the newly identified roAp stars as well as their G-band mean magnitudes and effective temperatures ($T_{\rm eff}$). 

\begin{table*}[h!]
\centering
\caption{The newly identified roAp stars with 200-s and 20-s cadence \tess\ data products: TIC\,312111544, TIC\,252881095, TIC\,46054683 and 49 Cam.}
\label{tab:obsall}
% Please add the following required packages to your document preamble:
% \usepackage{multirow}
%\resizebox{0.9\textwidth}{!}{%
%\small % or 
\footnotesize
\begin{tabular}{@{}ccccccccccccccc@{}}
\toprule
TIC                         & Cadence                 & G                      & $T_{\rm eff}$ [K]                     & Sector                      & $\bar t$ [BJD-2400000]       & $P_{\text{rot}}$ [d]     & $\sigma_{P_{\text{rot}}}$ [d] &                               & $\nu$ {[}d$^{-1}${]} & $\sigma_\nu$ {[}d$^{-1}${]} & $A$ {[}mmag{]} & $\sigma_{A}$ {[}mmag{]} & $\phi$ {[}rad{]} & $\sigma_\phi$ {[}rad{]} \\ \midrule
\multirow{12}{*}{312111544} & \multirow{12}{*}{200-s } & \multirow{12}{*}{9.23} & \multirow{12}{*}{7272$\pm110$}        & \multirow{3}{*}{61}         & \multirow{3}{*}{59975.97989} & \multirow{3}{*}{14.5925} & \multirow{3}{*}{0.0230}       & $\nu$-${\nu}_{\text{rot}}$    & 107.718              & 0.002                       & 0.091          & 0.008                   & 0.542            & 0.093                   \\
                            &                         &                        &                                       &                             &                              &                          &                               & $\nu$                         & 107.778              & 0.003                       & 0.069          & 0.008                   & 2.828            & 0.121                   \\
                            &                         &                        &                                       &                             &                              &                          &                               & $\nu$+${\nu}_{\text{rot}}$    & 107.854              & 0.002                       & 0.082          & 0.008                   & -2.046           & 0.102                   \\ \cmidrule(l){5-15} 
                            &                         &                        &                                       & \multirow{3}{*}{72}         & \multirow{3}{*}{60273.72543} & \multirow{3}{*}{14.0287} & \multirow{3}{*}{0.0226}       & $\nu$-${\nu}_{\text{rot}}$    & 107.730              & 0.004                       & 0.051          & 0.009                   & 1.890            & 0.170                   \\
                            &                         &                        &                                       &                             &                              &                          &                               & $\nu$                         & 107.783              & 0.002                       & 0.119          & 0.009                   & -1.436           & 0.073                   \\
                            &                         &                        &                                       &                             &                              &                          &                               & $\nu$+${\nu}_{\text{rot}}$    & 107.845              & 0.002                       & 0.114          & 0.009                   & 1.445            & 0.076                   \\ \cmidrule(l){5-15} 
                            &                         &                        &                                       & \multirow{3}{*}{88}         & \multirow{3}{*}{60705.04987} & \multirow{3}{*}{14.1230} & \multirow{3}{*}{0.0118}       & $\nu$-${\nu}_{\text{rot}}$    & 107.714              & 0.004                       & 0.060          & 0.009                   & 1.215            & 0.151                   \\
                            &                         &                        &                                       &                             &                              &                          &                               & $\nu$                         & 107.787              & 0.003                       & 0.064          & 0.009                   & -0.056           & 0.140                   \\
                            &                         &                        &                                       &                             &                              &                          &                               & $\nu$+${\nu}_{\text{rot}}$    & 107.855              & 0.003                       & 0.073          & 0.009                   & -2.018           & 0.121                   \\ \cmidrule(l){5-15} 
                            &                         &                        &                                       & \multirow{3}{*}{61, 72, 88} & \multirow{3}{*}{-}           & \multirow{3}{*}{14.7045} & \multirow{3}{*}{0.0004}       & $\nu$-${\nu}_{\text{rot}}$    & 107.710              & 0.002                       & 0.062          & 0.005                   & -                & -                       \\
                            &                         &                        &                                       &                             &                              &                          &                               & $\nu$                         & 107.778              & 0.001                       & 0.073          & 0.005                   & -                & -                       \\
                            &                         &                        &                                       &                             &                              &                          &                               & $\nu$+${\nu}_{\text{rot}}$    & 107.862              & 0.002                       & 0.071          & 0.005                   & -                & -                       \\ \midrule
\multirow{3}{*}{252881095}  & \multirow{3}{*}{200-s }  & \multirow{3}{*}{10.65} & \multirow{3}{*}{8547$\pm110$}         & 60                          & 59950.50933                  & 2.8865                   & 0.0020                        & $\nu$                         & 97.513               & 0.002                       & 0.293          & 0.021                   & 1.211            & 0.071                   \\ \cmidrule(l){5-15} 
                            &                         &                        &                                       & 73                          & 60296.34023                  & 2.8783                   & 0.0023                        & $\nu$                         & 97.517               & 0.002                       & 0.217          & 0.018                   & -2.873           & 0.082                   \\ \cmidrule(l){5-15} 
                            &                         &                        &                                       & 60, 73                      & -                            & 2.8917                   & 0.0015                        & $\nu$                         & 97.513               & 0.001                       & 0.293          & 0.014                   & -                & -                       \\ \midrule
\multirow{12}{*}{46054683}  & \multirow{12}{*}{200-s } & \multirow{12}{*}{9.75} & \multirow{12}{*}{7823$\pm110$}        & \multirow{4}{*}{71}         & \multirow{4}{*}{60247.31757} & \multirow{4}{*}{6.0074}  & \multirow{4}{*}{0.0033}       & $\nu_1$-${\nu}_{\text{rot}}$  & 101.875              & 0.001                       & 0.325          & 0.012                   & -0.612           & 0.038                   \\
                            &                         &                        &                                       &                             &                              &                          &                               & $\nu_1$                       & 102.047              & 0.001                       & 0.184          & 0.013                   & -2.860           & 0.070                   \\
                            &                         &                        &                                       &                             &                              &                          &                               & $\nu_1$+${\nu}_{\text{rot}}$  & 102.210              & 0.001                       & 0.230          & 0.013                   & 0.808            & 0.054                   \\
                            &                         &                        &                                       &                             &                              &                          &                               & $\nu_2$                       & 97.077               & 0.004                       & 0.062          & 0.013                   & -1.168           & 0.209                   \\ \cmidrule(l){5-15} 
                            &                         &                        &                                       & \multirow{4}{*}{72}         & \multirow{4}{*}{60273.53690} & \multirow{4}{*}{6.0098}  & \multirow{4}{*}{0.0019}       & $\nu_1$-${\nu}_{\text{rot}}$  & 101.777              & 0.001                       & 0.295          & 0.013                   & 2.626            & 0.044                   \\
                            &                         &                        &                                       &                             &                              &                          &                               & $\nu_1$                       & 101.950              & 0.003                       & 0.120          & 0.013                   & 1.190            & 0.110                   \\
                            &                         &                        &                                       &                             &                              &                          &                               & $\nu_1$+${\nu}_{\text{rot}}$  & 102.110              & 0.001                       & 0.200          & 0.013                   & -0.360           & 0.065                   \\
                            &                         &                        &                                       &                             &                              &                          &                               & $\nu_2$                       & -                    & -                           & -              & -                       & -                & -                       \\ \cmidrule(l){5-15} 
                            &                         &                        &                                       & \multirow{4}{*}{71, 72}     & \multirow{4}{*}{60260.37264} & \multirow{4}{*}{6.0015}  & \multirow{4}{*}{0.0019}       & $\nu_1$-${\nu}_{\text{rot}}$  & 101.821              & 0.001                       & 0.165          & 0.009                   & -2.189           & 0.055                   \\
                            &                         &                        &                                       &                             &                              &                          &                               & $\nu_1$                       & 101.977              & 0.001                       & 0.083          & 0.009                   & -3.101           & 0.105                   \\
                            &                         &                        &                                       &                             &                              &                          &                               & $\nu_1$+${\nu}_{\text{rot}}$  & 102.151              & 0.001                       & 0.101          & 0.009                   & 1.634            & 0.089                   \\
                            &                         &                        &                                       &                             &                              &                          &                               & $\nu_2$                       & 97.076               & 0.003                       & 0.036          & 0.009                   & 0.946            & 0.251                   \\ \midrule
\multirow{5}{*}{49 Cam}     & \multirow{5}{*}{20-s }   & \multirow{5}{*}{6.47}  & \multirow{5}{*}{7740$\pm$460$^{(a)}$} & \multirow{5}{*}{47}         & \multirow{5}{*}{59593.62261} & \multirow{5}{*}{4.2375}  & \multirow{5}{*}{0.0002}       & $\nu_1$-2${\nu}_{\text{rot}}$ & 122.109              & 0.002                       & 0.016          & 0.002                   & -2.918           & 0.120                   \\
                            &                         &                        &                                       &                             &                              &                          &                               & $\nu_1$                       & 122.576              & 0.003                       & 0.012          & 0.002                   & 2.018            & 0.166                   \\
                            &                         &                        &                                       &                             &                              &                          &                               & $\nu_1$+${\nu}_{\text{rot}}$  & 122.821              & 0.004                       & 0.010          & 0.002                   & 2.103            & 0.183                   \\
                            &                         &                        &                                       &                             &                              &                          &                               & $\nu_1$+2${\nu}_{\text{rot}}$ & 123.044              & 0.003                       & 0.011          & 0.002                   & -1.553           & 0.166                   \\
                            &                         &                        &                                       &                             &                              &                          &                               & $\nu_2$                       & 128.115              & 0.002                       & 0.018          & 0.002                   & 2.968            & 0.106                   \\ \bottomrule
\end{tabular}
%}
\tablefoot{In this table we show the results of non-linear least-squares fitting the light curves of the new 4 stars found using the significant peaks found in the DFT of each. We present their rotations periods, G-band mean magnitudes (G) from the \textit{Gaia} DR3 catalogue \citep{2023GaiaCollab}, effective temperatures ($T_{\rm eff}$), mean observation times ({$\bar t$}), ($P_{\text{rot}}$), frequencies ($\nu$), amplitudes ($A$), phases ($\phi$) and respective errors. For stars observed in more than one not consecutive sector, $\nu$, and the corresponding amplitude, denotes the adopted “true” pulsation frequency obtained from fitting the combined light curves, after resolving the correct alias. The listed $T_{\rm eff}$ are the effective temperatures from GSP-Phot Aeneas best library using BP/RP spectra reported in the \textit{Gaia} DR3 catalogue \cite{2023GaiaCollab}, unless otherwise stated. The uncertainty is the recommended value for \textit{Gaia} data as reported in \citealt{2023Fouesneau}.\\ \textbf{References.} a -- \citealt{2019Sikora_1}}
\end{table*}
\end{landscape}

\begin{table*}[h!]
\centering
\caption{TIC 312111544, TIC 46054683 and 49 Cam  - Oblique pulsator model.}
\resizebox{0.9\textwidth}{!}{%
\begin{tabular}{@{}ccccccccccc@{}}
\toprule
TIC                        & Cadence                & Sector                 & $t_0$ [BJD-2400000]          &                                   & $\nu$ [d$^{-1}$] & $\sigma_\nu$ [d$^{-1}$] & $A$ [mmag] & $\sigma_{A}$ [mmag] & $\phi$ [rad] & $\sigma_\phi$ [rad] \\ \midrule
\multirow{9}{*}{312111544} & \multirow{9}{*}{200-s } & \multirow{3}{*}{61}    & \multirow{3}{*}{59986.37757} & ${\nu}$-${\nu}_{\text{rot}}$      & 107.710          & 0.003                   & 0.084      & 0.008               & 1.233        & 0.100               \\
                           &                        &                        &                              & ${\nu}$                           & 107.778          & 0.003                   & 0.070      & 0.008               & 1.635        & 0.120               \\
                           &                        &                        &                              & ${\nu}$+${\nu}_{\text{rot}}$      & 107.847          & 0.003                   & 0.079      & 0.008               & 1.233        & 0.106               \\ \cmidrule(l){3-11} 
                           &                        & \multirow{3}{*}{72}    & \multirow{3}{*}{60281.60968} & ${\nu}$-${\nu}_{\text{rot}}$      & 107.712          & 0.002                   & 0.050      & 0.009               & 0.198        & 0.174               \\
                           &                        &                        &                              & ${\nu}$                           & 107.783          & 0.002                   & 0.119      & 0.009               & 0.239        & 0.073               \\
                           &                        &                        &                              & ${\nu}$+${\nu}_{\text{rot}}$      & 107.855          & 0.002                   & 0.104      & 0.009               & 0.198        & 0.084               \\ \cmidrule(l){3-11} 
                           &                        & \multirow{3}{*}{88}    & \multirow{3}{*}{60709.35858} & ${\nu}$-${\nu}_{\text{rot}}$      & 107.717          & 0.003                   & 0.061      & 0.009               & -2.457       & 0.148               \\
                           &                        &                        &                              & ${\nu}$                           & 107.787          & 0.003                   & 0.065      & 0.009               & -2.061       & 0.138               \\
                           &                        &                        &                              & ${\nu}$+${\nu}_{\text{rot}}$      & 107.858          & 0.003                   & 0.073      & 0.009               & -2.457       & 0.123               \\ \midrule
\multirow{9}{*}{46054683}  & \multirow{9}{*}{200-s } & \multirow{3}{*}{71}    & \multirow{3}{*}{60249.71119} & ${\nu}$-${\nu}_{\text{rot}}$      & 101.880          & 0.001                   & 0.306      & 0.013               & -1.469       & 0.041               \\
                           &                        &                        &                              & ${\nu}$                           & 102.047          & 0.001                   & 0.196      & 0.013               & -1.243       & 0.066               \\
                           &                        &                        &                              & ${\nu}$+${\nu}_{\text{rot}}$      & 102.213          & 0.001                   & 0.228      & 0.013               & -1.469       & 0.056               \\ \cmidrule(l){3-11} 
                           &                        & \multirow{3}{*}{72}    & \multirow{3}{*}{60273.79837} & ${\nu}$-${\nu}_{\text{rot}}$      & 101.783          & 0.003                   & 0.279      & 0.013               & 0.055        & 0.047               \\
                           &                        &                        &                              & ${\nu}$                           & 101.949          & 0.003                   & 0.074      & 0.013               & 0.199        & 0.180               \\
                           &                        &                        &                              & ${\nu}$+${\nu}_{\text{rot}}$      & 102.116          & 0.003                   & 0.188      & 0.013               & 0.055        & 0.071               \\ \cmidrule(l){3-11} 
                           &                        & \multirow{3}{*}{71,72} & \multirow{3}{*}{60261.42557} & ${\nu}$-${\nu}_{\text{rot}}$      & 101.810          & 0.001                   & 0.142      & 0.009               & 0.528        & 0.064               \\
                           &                        &                        &                              & ${\nu}$                           & 101.977          & 0.001                   & 0.065      & 0.009               & 0.890        & 0.139               \\
                           &                        &                        &                              & ${\nu}$+${\nu}_{\text{rot}}$      & 102.144          & 0.001                   & 0.085      & 0.009               & 0.528        & 0.107               \\ \midrule
\multirow{5}{*}{49 Cam}    & \multirow{5}{*}{20-s }  & \multirow{5}{*}{47}    & \multirow{5}{*}{59594.45976} & ${\nu}_{1}$-2${\nu}_{\text{rot}}$ & 122.111          & 0.003                   & 0.016      & 0.002               & -1.506       & 0.121               \\
                           &                        &                        &                              & ${\nu}_{1}$-${\nu}_{\text{rot}}$  & 122.343          & 0.003                   & 0.001      & 0.002               & 0.403        & 1.700               \\
                           &                        &                        &                              & ${\nu}_{1}$                       & 122.576          & 0.003                   & 0.012      & 0.002               & -0.403       & 0.161               \\
                           &                        &                        &                              & ${\nu}_{1}$+${\nu}_{\text{rot}}$  & 122.809          & 0.003                   & 0.008      & 0.002               & 1.004        & 0.224               \\
                           &                        &                        &                              & ${\nu}_{1}$+2${\nu}_{\text{rot}}$ & 123.042          & 0.003                   & 0.012      & 0.002               & -1.506       & 0.157               \\ \bottomrule
\end{tabular}
}
\tablefoot{Frequencies ($\nu$), amplitudes ($A$), phases ($\phi$) and respective errors, found by forcing the sidelobes to be separated from $\nu$ by exactly $\pm\nu_{\text{rot}}$ and $\pm2\nu_{\text{rot}}$ and fitting the light curve of each Sector at the time of pulsation maximum $t_0$.}
\label{tab:force_fit}
\end{table*}

%\clearpage

\section{LAMOST spectra}
\label{spectra}

In this section we show the LAMOST spectra of the newly identified roAp stars (TIC\,46054683, TIC\,252881095, and TIC\,312111544) as well as standard stars of similar spectral type to highlight the spectral peculiarities. Continuum subtraction was done using a similar approach to that used in \citet{2022Lazarz}.

\begin{figure*}[h!]
    \centering
    \includegraphics[width=\textwidth]{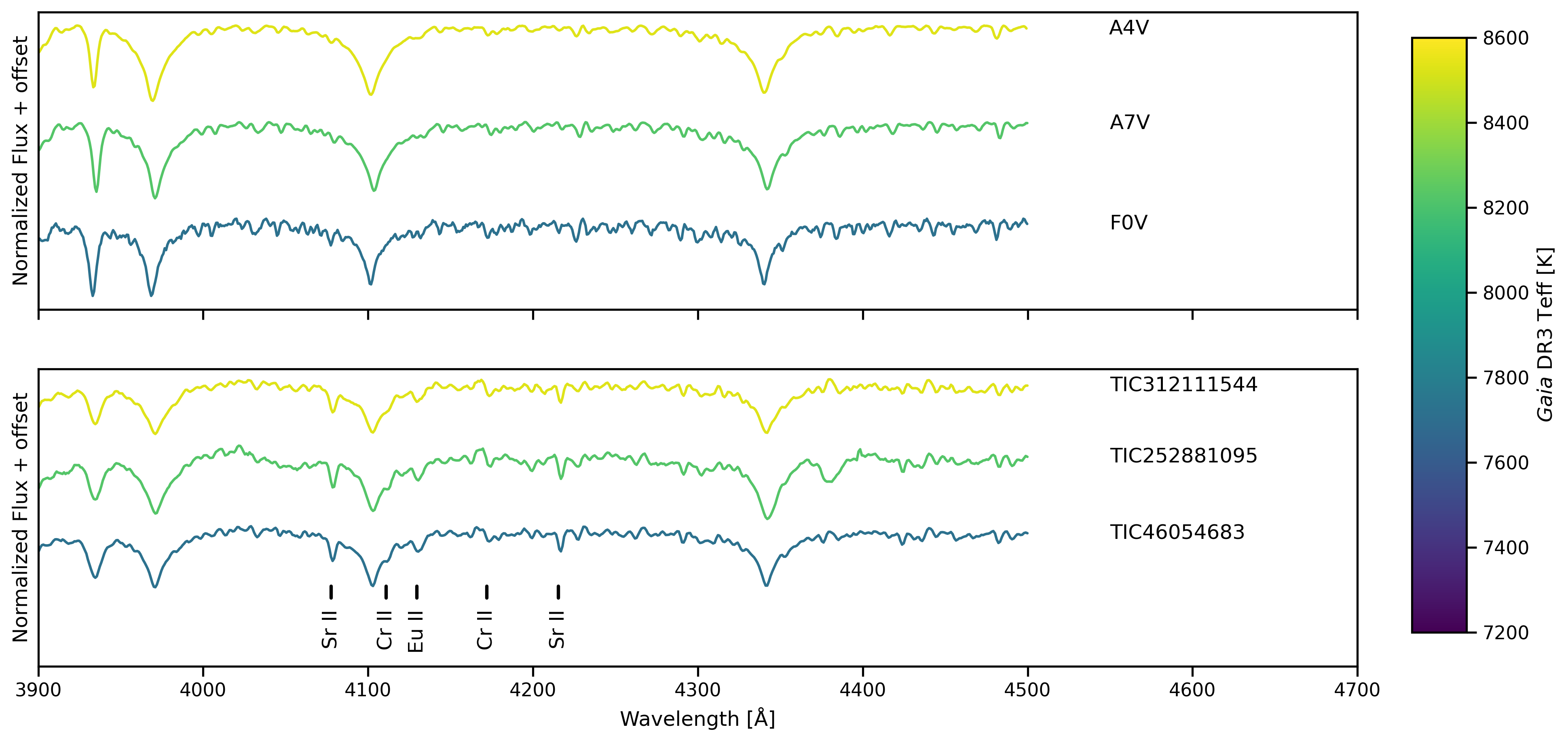}
    \caption{\textit{Top panel:} Stellar spectra of three standard stars of spectral type A4V, A7V and F0V with temperatures similar to those of TIC\,46054683, TIC\,252881095 and TIC\,312111544 respectively. \textit{Bottom panel:} LAMOST low-resolution spectra of TIC\,46054683, TIC\,252881095 and TIC\,312111544 in the blue region coloured according to the \textit{Gaia} DR3 \texttt{teff\_gspphot} for each star.}
    \label{fig:lamost_blue}
\end{figure*}
%\end{landscape}

\clearpage
\twocolumn
\section{Additional figures}
\label{gaiafigs}

\begin{figure}[h!]
    \centering
    \includegraphics[scale=0.4]{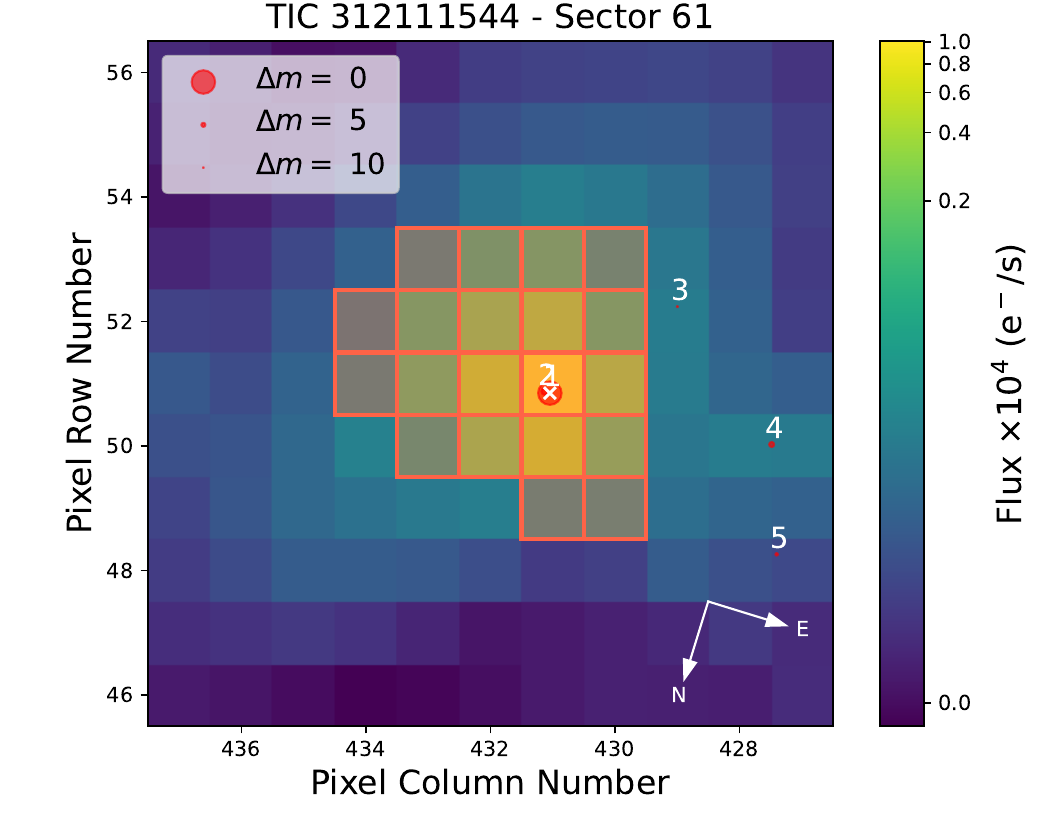}
    \caption{As Figure \ref{fig:binary} for TIC\,312111544 observed in Sector 61.}
    \label{fig:TPF_312111544}
\end{figure}

\begin{figure}[h!]
    \centering
    \includegraphics[scale=0.4]{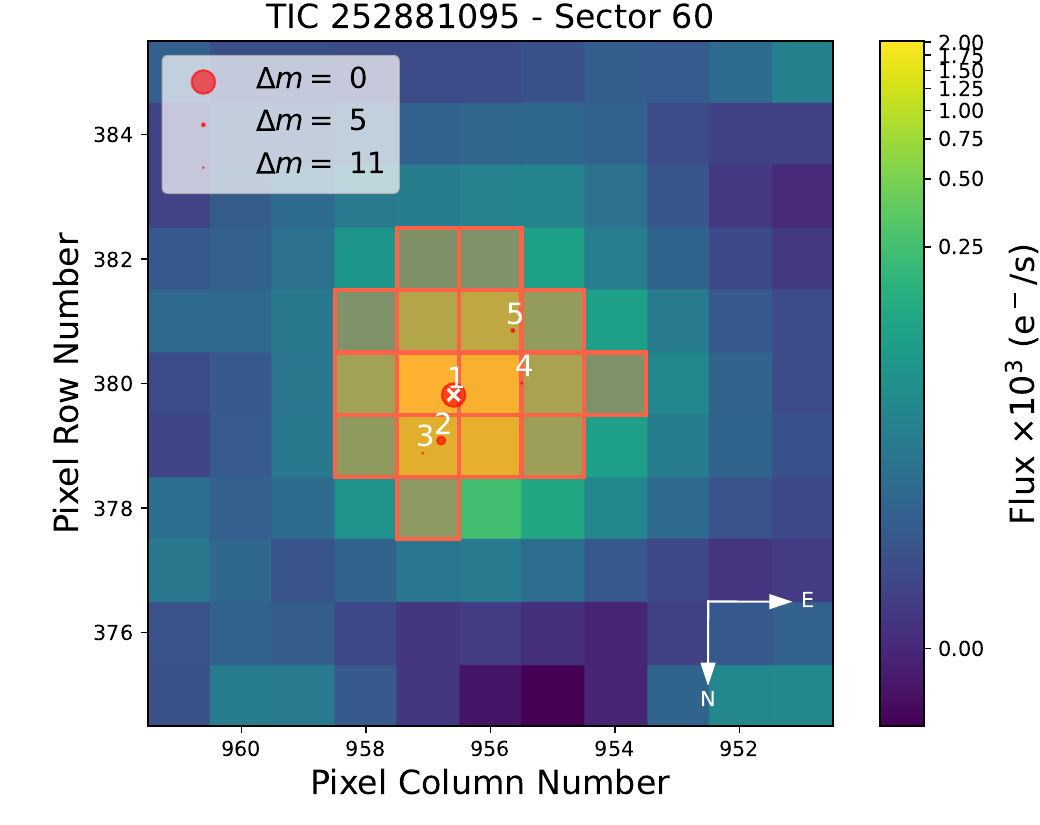}
    \caption{As Figure \ref{fig:binary} but for TIC\,252881095 observed in Sector 60.}
    \label{fig:TPF_252881095}
\end{figure}

\section{Inference of binary system parameters}
\label{MCMC}

In this section we describe the methods we use to infer the parameters of Equation~\ref{binary_eq} and estimate the physical properties of binary systems from \tess\ data. It is worth noting that Equation~\ref{binary_eq} describes a circular orbit. We use nested sampling (\texttt{dynesty}) to explore the full posterior over sinusoid parameters ($A_{\tau}$, $C$, $\varphi$ and $P_{\text{orb}}$) and to identify the solutions that are most likely to fit the observed $\tau(t)$ within a 68 \% credible interval. This problem is highly aliased for two main reasons: i) \tess\ data are rarely consecutive, so the temporal separation between observed sectors introduces multiple, nearly degenerate solutions for $P_{\text{orb}}$; {\sc{ii}}) each \tess\ sector spans only $\sim$ 27\,d, and roAp stars typically rotate slowly which yields few one-rotation light-curve segments and therefore few independent phase/time-delay measurements, as calculated from the pulsation averaged over one stellar rotation cycle, with which to fit Equation~\ref{binary_eq}. Nested sampling is well suited to deal with such a problem. Essentially, the parameter space is explored using \textit{live points}, i.e. $N_{\text{live}}$ solutions of Equation~\ref{binary_eq}, each parameter randomly sampled from its given prior. It then iteratively removes the lowest likelihood solutions and draws a new one, forcing the solution to have higher likelihood than the one it removed. The sequence of discarded/replaced points yields posterior samples, and summing the likelihood over the shrinking prior volume yields the marginal likelihood for model comparison.

We choose to use the following log-likelihood function,

\begin{equation}
\ln \mathcal{L} \;=\; -\tfrac12 \sum_{i}
\left[
\ln\!\big(2\pi s_i^2\big) \;+\; \frac{\big(\tau_i - \tau(t_i)\big)^2}{s_i^2}
\right],
\end{equation}
valid under the assumption that the $\tau(t_i)$ measurements have approximately independent, Gaussian errors $s_i=\sigma_i^2 \;+\; \sigma_{\rm jit}^2 $. The extra “jitter” term $\sigma_{\rm jit}>0$ accounts for mild error underestimation and low-level systematics. We also subtract a global mean from $\tau(t)$ to centre it around zero and we break the sign degeneracy, $A_{\tau}\sin\theta = -A_{\tau}\sin(\theta+\pi)$, by exploring the parameter space only for $A_{\tau}\ge 0$. As ($A_{\tau}$,\,$\varphi$) and (-$A_{\tau}$,\,$\varphi+\pi$) describe the same curve, forcing $A_{\tau}\ge 0$ ensures we count each sinusoid only once. We adopt broad, uninformative priors for each of the parameters we fit. For \(A_{\tau}\), \(C\), \(P_{\rm orb}\), and \(\sigma_{\rm jit}\) we choose log–uniform (Jeffreys–like) priors and for $\phi$ we choose a uniform prior. We define the prior bounds as follows: 

\begin{itemize}
  \item {Amplitude $A_{\tau}\geq0$:} log-uniform on $[A_{\tau}^{\min},A_{\tau}^{\max}]=[10^{-4},1000]$ s;
  \item {Phase $\varphi$:} uniform on $[0,2\pi]$;
  \item {Constant $C$\, ($C\geq0$ for numerical convenience):} log-uniform in $[C_{\min},C_{\max}]=[0,100]$ s;
  \item {Period $P_{\rm orb}$:} log-uniform on $[P_{\min},P_{\max}]=[1,1000]$ d;
  \item {Jitter $\sigma_{\rm jit}>0$:} log-uniform on $[\sigma_{\min},\sigma_{\max}]=[10^{-6},100]$ s.
\end{itemize}

To validate our method we apply it to $\tau(t)$ obtained from \tess\ 120-s cadence data of TIC\,33601621 in Sectors 6, 37 and 86. This is the only known roAp star in a spectroscopic binary system \citep{2019Holdsworth}. The results can be seen in Figure \ref{fig:known_bin_mcmc}. Using this method we estimate $P_{\text{orb}}\sim$ 90.1 days, with a 68\% credible interval of [72.1,173] days. These results are compatible with the orbital parameters obtained from spectroscopy as TIC\,33601621 is part of a binary system with an eccentric orbit ($e$=0.317) and an orbital period of $P_{\text{orb}}$=93.266 days.

\begin{figure}[h!]
    \centering
    \includegraphics[width=\columnwidth]{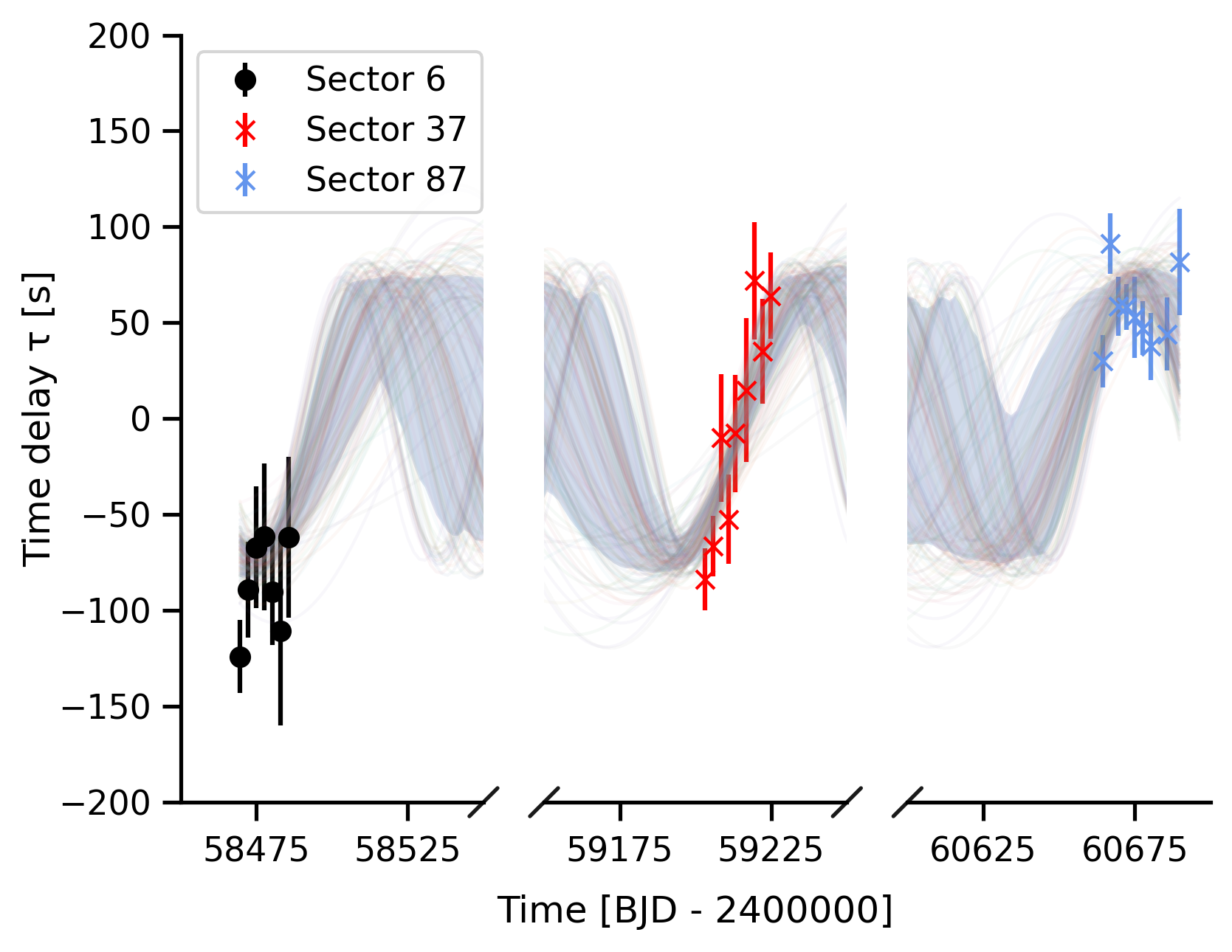}
    \caption{Most likely solutions of Equation~\ref{binary_eq} to fit the $\tau$ variability over time observed for TIC\,33601621.}
    \label{fig:known_bin_mcmc}
\end{figure}

\section{Impact of cadence and brightness on the detectability of roAp pulsations}
\label{20secbin}

In this section, we study how the photometric cadence and target brightness impact the SNR of roAp pulsation detections in \tess\ light curves. We use 20-s SPOC light curves for two previously known roAp stars, listed in Table~\ref{roAp_test}, and artificially degrade each light curve to longer effective cadences (20 to 200-sin steps of 20-s). 

\begin{table}[h!]
\centering
\caption{List of previously known roAp stars used to test the detectability of roAp pulsations.}
%\resizebox{\columnwidth}{!}{%
\footnotesize
\begin{tabular}{@{}ccccc@{}}
\toprule
TIC       & G$_{\text{mag}}$ [mag] & $\nu$ [d$^{-1}$] & $A$ [mmag] & Sector \\ \midrule
280198016 & 6.2                    & 123.0            & 0.81     & 90     \\
273777265 & 13.7                   & 129.0            & 1.06     & 55     \\ \bottomrule
\end{tabular}
%}
\tablefoot{For each target we list the TIC ID, the \textit{Gaia} G band mean magnitude, G, the frequency, $\nu$, and amplitude, $A$, of the dominant pulsation frequency determined by non-linear least-squares fitting the 20-s cadence light curve observed in the listed Sector.}
\label{roAp_test}
\end{table}

\begin{figure}[h!]
    \centering
    \includegraphics[width=\columnwidth]{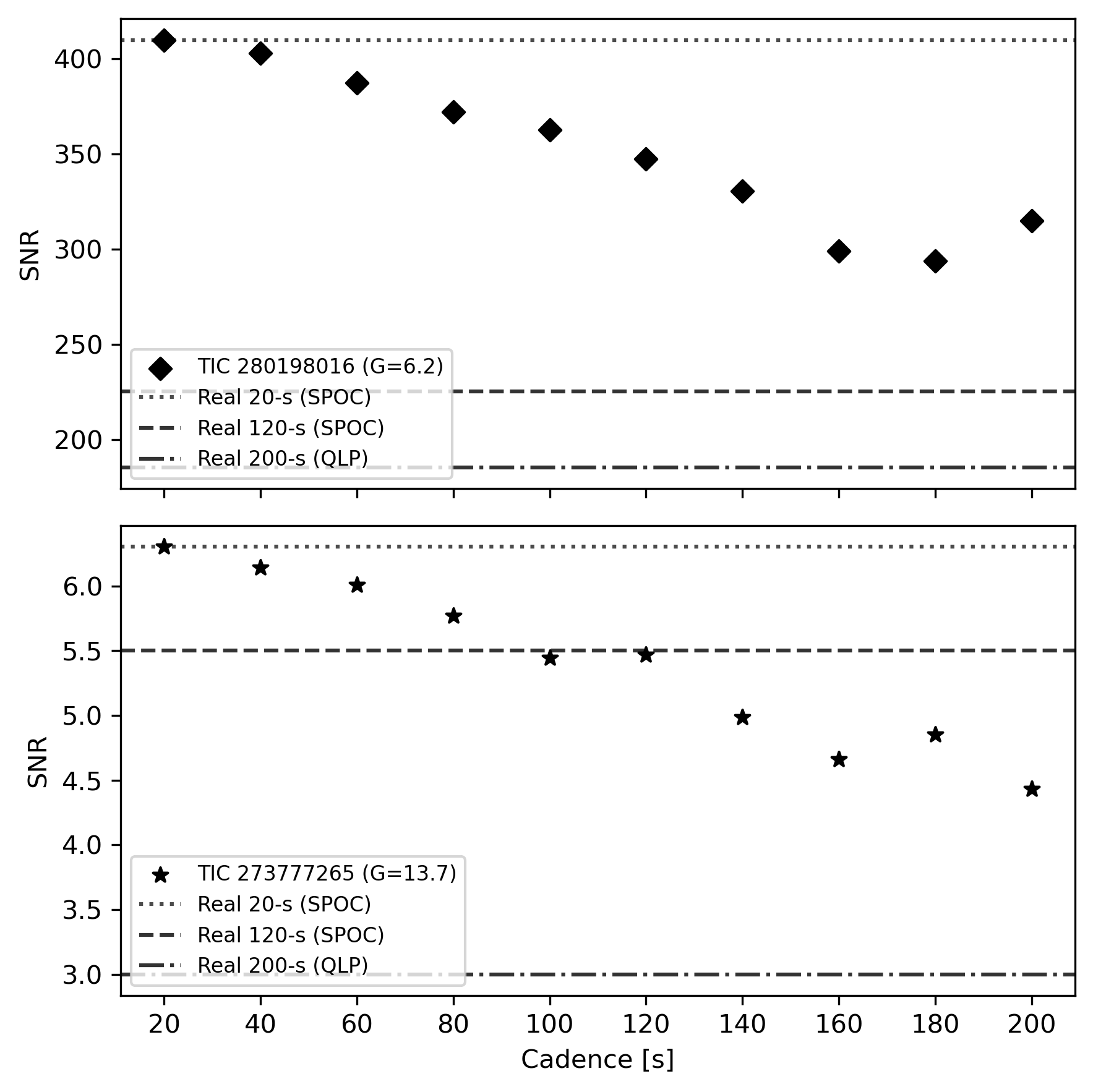}
    \caption{Variation of the SNR of the dominant roAp pulsation as a function of effective cadence for TIC\,280198016 (top panel; Sector 90) and TIC\,273777265 (bottom panel; Sector 55). Filled symbols show the SNR measured from the DFT computed from light curves that were artificially degraded from the original 20-s SPOC data to longer cadences. Horizontal lines mark the SNR measured directly from original \tess\ light curves. The dotted and dashed lines indicate the 20-s and 120-s SPOC products for Sectors 90 (TIC\,280198016) and 55 (TIC\,273777265). The dash–dotted line shows the SNR computed from 200-s QLP light curves Sector 90 (TIC\,280198016) and Sector 82 (TIC\,273777265).}
    \label{fig:pts}
\end{figure}

\begin{figure}[h!]
    \centering
    \includegraphics[width=\columnwidth]{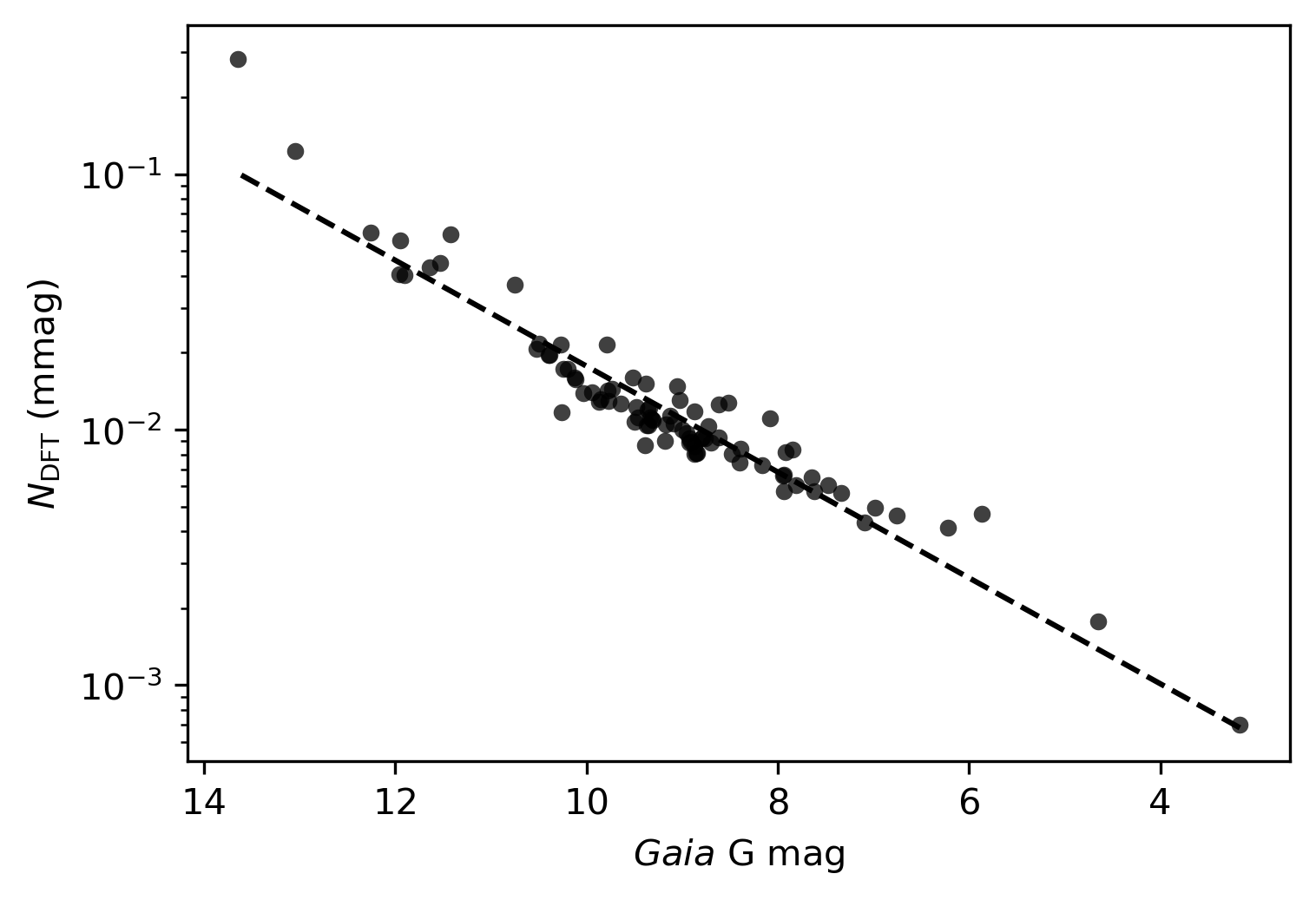}
    \caption{Median high-frequency noise level in the DFT, $N_{\rm DFT}$, measured from 200-s QLP light curves of previously known roAp stars as a function of \textit{Gaia} G magnitude. The dashed line shows a linear fit in $\log_{10}(N_{\rm DFT})$ as a function of G.}
    \label{fig:mags}
\end{figure}

\citet{2013PhDT.......638M} described mathematically how finite integration time reduces the measured Fourier amplitudes of high-frequency pulsations. Longer exposures act as an averaging filter, such that the observed amplitude in the DFT is attenuated relative to the intrinsic stellar signal. Because our goal is to investigate how detectability changes as the noise level increases with cadence (and with target brightness), we treat the intrinsic mode amplitude as fixed for each star. We note that, for a given cadence, higher-frequency pulsations suffer stronger attenuation because the signal is averaged over a larger fraction of the pulsation cycle. This effect becomes increasingly relevant for modes approaching the Nyquist frequency. Our test stars have similar pulsation frequencies (Table~\ref{roAp_test}), so this does not affect the relative comparison presented here.

For each target, we adopt the dominant (highest-amplitude) pulsation mode as the reference signal. To minimise integration-time attenuation, we determine the reference frequency and amplitude by least-squares fitting the 20-s light curve of each target (Table~\ref{roAp_test}). We then degrade the 20-s light curve to longer effective cadences and compute the DFT for each degraded data set. We correct the amplitudes in each degraded amplitude spectrum for the expected integration-time attenuation using the sinc response appropriate for that effective cadence. Finally, we compute the SNR of the dominant pulsation frequency at each cadence using the fixed reference amplitude and the (attenuation-corrected) local noise level measured in the amplitude spectrum. With this approach, changes in SNR with cadence primarily reflect changes in the noise level, rather than differences in the intrinsic mode amplitude.

The results in Figure \ref{fig:pts} suggest that the SNR of the dominant pulsation frequencies of TIC\,273777265 and TIC\,280198016 decreases as the time interval between measurements in the light curve increases. For TIC\,273777265, the SNR drops from 6.30 in the 20-s light curve to 4.43 in the 200-s degraded light curve, corresponding to a decrease of $\sim$29.7\%. TIC\,280198016 shows a similar cadence dependence, with the SNR decreasing from 410.01 (20-s) to 315.09 (200-s), i.e. a $\sim$23.2\% reduction.

Comparing the original SPOC products, we find that the SNR in the 120-s light curves is lower than in the 20-s data by $\sim$12.7\% for TIC\,273777265 and $\sim$45.0\% for TIC\,280198016. For TIC\,273777265, the SNR measured from the original 120-s SPOC light curve is consistent with the value predicted from the artificially degraded 20-s light curve to within $\sim$0.6\%. At 200-s, the SNR measured from the 200-s QLP light curve is $\sim$32.3\% lower than the value predicted from the artificially degraded 20-s light curve. This reduces the detection to SNR$\approx$3 and therefore below our adopted detectability threshold (SNR$\approx$4.5), such that the dominant pulsation is no longer recovered in the 200-s QLP data of TIC\,273777265.

Figure \ref{fig:pts} also illustrates how brightness influences detectability. TIC\,280198016 (\textit{Gaia} $G=6.2$) has a higher SNR than the fainter TIC\,273777265 (\textit{Gaia} $G=13.7$) at all cadences, consistent with the higher photometric noise expected for fainter targets.

Our results are consistent with \citet{2022Huber}, who showed that 20-s SPOC light curves binned to an effective 120-s cadence typically achieve better photometric precision than the native SPOC 120-s products, and that this discrepancy is strongest for bright targets. The authors mainly attribute this effect to the different treatment of cosmic rays in the two data products. In the SPOC 120-s pipeline, a single cosmic-ray hit can cause the rejection of the entire 120-s exposure, leading to the loss of a substantial fraction of measurements ($\sim$20\%; \citealt{2022Huber}). In contrast, for the 20-s cadence data, a cosmic ray generally impacts only a single 20-s exposure, so the same event removes less information when the 20-s data are binned to 120-s . This is more severe for brighter stars because their intrinsic photon noise is low, so the loss of exposures impacts precision more negatively than it does for fainter targets whose noise is already dominated by photon and background contributions \citep{2022Huber}. Our two test cases reflect such behaviour well: TIC\,280198016 (G=6.2) shows a larger discrepancy than TIC\,273777265 (G=13.7) when comparing the SNR expected from 20-s cadence data binned to 120-s versus the SNR measured from the original 120-s SPOC data.

For TIC\,280198016, the 120-s SPOC SNR is $\sim$35.1\% lower than the value predicted from the artificially degraded 20-s light curve, and the 200-s QLP SNR is $\sim$41.2\% lower than the value predicted from the artificially degraded 20-s light curve. This additional loss of SNR in the QLP products relative to the value predicted from the artificially degraded 20-s light curve most likely reflects differences in light-curve extraction and processing between the QLP and SPOC pipelines. The QLP light curves are extracted from FFIs using simple calibration and aperture-based extraction, in comparison to the SPOC products that undergo instrumental corrections and detrending. In practice, these differences mainly act to raise the high-frequency noise level in the QLP amplitude spectra, so the impact on detectability is also expected to be brightness dependent. 

To quantify this dependency, we measured the high-frequency noise level in the amplitude spectrum, $N_{\rm DFT}$, from 200-s QLP light curves for all the previously known roAp stars observed by \tess\ \citep{Cycle_1_2021,Cycle_2_2024}. For each star we compute a median $N_{\rm DFT}$ from the available QLP light curves to mitigate sector-to-sector variations. As shown in Figure~\ref{fig:mags}, the DFT noise level decreases toward brighter \textit{Gaia} G magnitudes. Thus low-amplitude, high-frequency pulsations become harder to recover with 200-s QLP data for fainter targets. This magnitude dependence supports our conclusion that the faintness of LAMOST targets is most likely related to why we are unable to recover the roAp incidence among Ap stars relative to the $\sim$5.5\% reported in \citet{Cycle_2_2024}.

Overall, these results show that 200-s cadence data provide a valuable resource for detecting roAp pulsators, particularly for brighter targets. However, short-cadence data remain preferable for detailed pulsation characterisation, including the recovery of low-amplitude modes and rotationally split sidelobe frequencies. As such, we highlight the importance of acquiring \tess\ and PLATO data at the shortest possible cadence for comprehensive mode identification in roAp stars.

\end{appendix}
\end{document}